\documentclass[reqno]{amsart}

\usepackage{amsthm,amsmath,amsfonts,amssymb,epsfig,graphicx,latexsym,float,color}
\usepackage{pgfplots}
\usepackage{subfigure}
\usepackage{multirow}
\usetikzlibrary{plotmarks}
\usetikzlibrary{external}
\newtheorem{definition}{Definition}

\newtheorem{remark}[definition]{Remark}

\newtheorem{system}{System}

\renewcommand{\theequation}{\arabic{section}.\arabic{equation}}
\renewcommand{\thetable}{\arabic{section}.\arabic{table}}
\renewcommand{\thefigure}{\arabic{section}.\arabic{figure}}

\newcommand{\tn}[1]{\textnormal{#1}}
\newcommand{\bs}[1]{\boldsymbol{#1}}

\newcommand{\ddz}{\frac{\mathrm{d}}{\mathrm{d} z}}
\newcommand{\deriv}[1]{\frac{\mathrm{d} #1}{\mathrm{d} z}}

\newcommand{\closureFr}{\mathcal{U}_r}
\newcommand{\closureFz}{\mathcal{U}_z}
\newcommand{\closureFzTilde}{\tilde{\mathcal{U}}_z}

\newcommand{\abk}{\mathcal{R}} 

\begin{document}

\title[]{On flow-enhanced crystallization in fiber spinning: Asymptotically justified boundary conditions for numerics of a stiff viscoelastic two-phase model}
\author[Ettm\"uller et al.]{Manuel Ettm\"uller$^{1}$}
\author[]{Walter Arne$^{1}$}
\author[]{Nicole Marheineke$^{2}$}
\author[]{Raimund Wegener$^{1}$}

\date{\today\\
$^1$ Fraunhofer ITWM, Fraunhofer Platz 1, D-67663 Kaiserslautern, Germany\\
$^2$ Universit\"at Trier, Lehrstuhl Modellierung und Numerik, Universit\"atsring 15, D-54296 Trier, Germany}

\begin{abstract}
For flow-enhanced crystallization in fiber spinning, the viscoelastic two-phase fiber models by Doufas et al.\ (J.\ Non-Newton.\ Fluid Mech., 2000) and Shrikhande et al.\ (J.\ Appl.\ Polym. Sci., 2006) are state of the art. However, the boundary conditions associated to the onset of crystallization are still under discussion, as their choice might cause artificial boundary layers and numerical difficulties. In this paper we show that the model class of ordinary differential equations is singularly perturbed in a small parameter belonging to the semi-crystalline relaxation time and derive asymptotically justified boundary conditions. Their effect on the overall solution behavior is restricted to a small region near the onset of crystallization. But their impact on the performance of the numerical solvers is huge, since artificial layering, ambiguities and parameter tunings are avoided. The numerics becomes fast and robust and opens the field for simulation-based process design and material optimization.
\end{abstract}

\maketitle

{\sc AMS-Classification.} 76-XX, 34Bxx, 34E15, 65L10

\setcounter{equation}{0} \setcounter{figure}{0} \setcounter{table}{0}
\section{Introduction and model class}\label{sec_introduction}
In the technical textile industry melt spinning is one of the most important processes for manufacturing long slender polymer fibers. Spinning semi-crystalline polymers like polyamide (Nylon) or polyethylene terephthalate (PET), crystallization occurs along the spinline which crucially affects the properties of the produced fabrics. Thus, for simulation-based process design and material optimization the flow-enhanced crystallization has to be taken into account in the mathematical model.

The trending model comes from Doufas et al., \cite{doufas_simulation_2001, doufas_simulation_2000-1, doufas_simulation_2000}. It is a stationary one-dimensional viscoelastic two-phase fiber model that describes the two phases occurring in melt spinning along the spinline, -- the amorphous phase and the semi-crystalline phase --, by separate constitutive equations. The equations are coupled via balance laws and an evolution equation for the degree of crystallization. Comparisons with experimental data show that the model is suitable for successfully simulating melt spinning of various polymers under low and high speed spinning conditions. The boundary value problem of ordinary differential equations, however, creates artificial discontinuities in the solution at the point of crystallization onset and at the point of complete crystallization. 
Thus, Shrikhande et al.\ \cite{shrikhande_modified_2006} proposed a modified model by introducing a crystallization rate that depends on the stored free energy of the molten phase, including a respective additional evolution equation for the free energy and exchanging the boundary conditions that address the crystallization onset. An other model variant can be found in \cite{dhadwal_paper_2011} where the original system of equations is supplemented by modified boundary conditions that are based on a force balance at the point of crystallization onset. In all mentioned works the choice of the boundary conditions addressing the onset of crystallization is heuristically motivated. As it crucially affects the solution behavior concerning the possible rising of artificial boundary layers or even discontinuities, it has an immense impact on the performance of the numerical solvers. 

Aiming for simulation and optimization of industrial spinning processes with thousands of fibers interacting with a surrounding airflow, a robust and efficient numerics is required which goes with an analytical understanding of the underlying boundary value problem. In this paper we show that the two-phase fiber models by Doufas et al.\ and Shrikhande et al.\ are mathematically of similar structure and can be treated in a common model class. Our investigations reveal that the underlying system of ordinary differential equations is perturbed by a small model parameter $\delta$ associated to the semi-crystalline relaxation time. In the limit $\delta=0$, the system degenerates to a system of differential-algebraic equations. The algebraic relations prescribe the orientational tensor component and the velocity changes which are associated with the onset of crystallization. We propose a regular perturbation and use asymptotically derived relations to provide suitable boundary conditions for the two-phase fiber model.  They are consistent to all models variants from literature and yield boundary value problems with smooth solutions. In the numerical simulations the effects of our asymptotically justified boundary conditions on the overall solution behaviors -- that are known from  \cite{doufas_simulation_2000} and \cite{shrikhande_modified_2006} -- seem to be marginal, as they are restricted to very small regions near the point of crystallization onset. But the improvement of the performance of the numerical solvers is huge, since artificial boundary layers, ambiguities and parameter tunings are avoided. The numerics becomes fast and robust. We comment on the advantages of our collocation-continuation method over the shooting approach used in the cited literature.

The paper is structured as follows.
The focus is on the well-established two-phase fiber models by Doufas et al.\ and Shrikhande et al.\ that we briefly summarize in the following introductory subsection. In Section~\ref{sec_asymptotic} we analyze the mathematical structure of the underlying equation system in a non-dimensional form and derive asymptotically justified boundary conditions, which enable the formulation of regularly perturbed model problems. The impact of our modifications on the numerical solvers and the simulation results are discussed for melt spinning of Nylon-66 in Section~\ref{sec:numerics}. The appendix provides further details to the asymptotic derivation, the used closure models (such as for material properties, heat transfer, air drag), the test case setup and our numerical treatment.

\subsection{Two-phase fiber model class for flow-enhanced crystallization}
The two-phase fiber model originally proposed by Doufas et al.\ in \cite{doufas_simulation_2000-1} and modified by Shrikhande et al.\ in \cite{shrikhande_modified_2006} assumes a stationary uniaxial spinning setup with a circular-shaped viscoelastic fiber under gravity immersed in an airflow. It combines radially averaged balance laws for mass, momentum and energy with constitutive equations for the amorphous melt phase and the semi-crystalline phase that are coupled via an evolution equation for the degree of crystallization. The unknown variables are the cross-sectional area $A$, the axial velocity $v_z$, the temperature $T$, the conformation tensor components $c_{zz}$ and $c_{rr}$ related to the amorphous phase, the orientational tensor component $S_{zz}$ related to the semi-crystalline phase as well as the crystallinity $x$ along the spinline $z\in[0,L]$. In \cite{shrikhande_modified_2006} the stored free energy $a$ of the molten phase enters as additional unknown.

The model system is given by
\begin{subequations}\label{eq:model}
\begin{align}
\ddz (v_z \rho A)																&= 0, \\
v_z \rho A \deriv{v_z}															&= \ddz \Big( A \left( \tau_{zz}-\tau_{rr} \right) \Big) - \pi B \mu_\mathrm{a} \left( v_z - v_\mathrm{a}^\parallel \right) + \rho A g+ \frac{\pi}{2} \gamma \deriv{D}, \\
v_z \rho A C_\mathrm{p}\deriv{T}												&= - \pi D h (T-T_\mathrm{a}) + A \left( \tau_{zz}-\tau_{rr} \right) \deriv{v_z}\ + \rho A \Phi_{\infty}\Delta H_\mathrm{f} v_z \deriv{x}, \\
\lambda_\mathrm{am} \left( v_z \deriv{c_{zz}} - 2c_{zz}\deriv{v_z} \right) 		&= - \bigg(1-\alpha + \frac{ \alpha}{\zeta} \frac{c_{zz}}{1-x} \bigg) \bigg(c_{zz} - \zeta (1-x) \bigg), \\
\lambda_\mathrm{am} \left( v_z \deriv{c_{rr}} + c_{rr}\deriv{v_z} \right)		&= - \bigg(1-\alpha + \frac{ \alpha}{\zeta} \frac{c_{rr}}{1-x} \bigg) \bigg(c_{rr} - \zeta (1-x) \bigg), \\
\lambda_\mathrm{sc} \left( v_z \deriv{S_{zz}} - 2 S_{zz}\deriv{v_z} \right)		&=  - \sigma S_{zz} + \lambda_\mathrm{sc} \left( \frac{2}{3} -2 \closureFz(S_{zz}) \right) \deriv{v_z}, \\
v_z \deriv{x} 																	&= K(1-x) 
\end{align}
with 
\begin{align*}
\tau_{zz}	&= G \left(\frac{1}{\zeta} \frac{c_{zz}}{1-x} -1 + 3 S_{zz} + 6 \lambda_\mathrm{sc} \closureFz(S_{zz}) \deriv{v_z} \right),\\ \qquad
\tau_{rr}	&= G \left(\frac{1}{\zeta} \frac{c_{rr}}{1-x} -1 - \frac{3}{2} S_{zz} + 6 \lambda_\mathrm{sc} \closureFr(S_{zz}) \deriv{v_z} \right), \\
\lambda_\mathrm{am} &= \lambda(1-x)^2,\qquad   \lambda_\mathrm{sc} = \delta \lambda\exp(Fx),\qquad   \lambda = \frac{\mu(T)}{G},
\end{align*}
where $\tau_{zz}$, $\tau_{rr}$ are the stress tensor components and $\lambda_\mathrm{am}$ and $\lambda_\mathrm{sc}$ the amorphous and semi-crystalline relaxation times, respectively. 
The crystallization rate $K$ is modeled by Doufas et al.\ \cite{doufas_simulation_2000-1} as 
\begin{align*}
K(T,\tau_{zz}, \tau_{rr}) &= K_\mathrm{max} \exp\bigg(-4 \ln(2) \frac{(T-T_\mathrm{max})^2}{(\Delta T)^2} + \frac{\xi}{G} \left(\tau_{zz} + 2 \tau_{rr}\right) \bigg)
\end{align*} 
and by Shrikhande et al.\ \cite{shrikhande_modified_2006} as
\begin{align}
\lambda_\mathrm{am} v_z \deriv{a}												&= -a + \lambda_\mathrm{am} \frac{G}{\zeta} \frac{(c_{zz} - c_{rr})}{1-x} \deriv{v_z},\\
K(T,a) &= K_\mathrm{max} \exp\bigg(-4 \ln(2) \frac{(T-T_\mathrm{max})^2}{(\Delta T)^2} + 2 \frac{\xi}{G} a \bigg). \nonumber
\end{align}
\end{subequations}

In \eqref{eq:model} fiber density $\rho(T,x)$, dynamic viscosity $\mu(T)$, specific heat capacity $C_\mathrm{p}(T,x)$ and specific latent heat of crystallization $\Delta H_\text{f} (T,x)$ might be chosen as temperature and crystallinity dependent functions. The aerodynamic forces on the fiber dynamics are formulated in terms of the Bingham function $B(v_z,D,T)$, the viscosity $\mu_\mathrm{a}(T)$ and tangential velocity component $v_\mathrm{a}^\parallel$ of the air. The heat transfer between fiber and air is prescribed by means of heat transfer coefficient $h(v_z,D,T)$ and air temperature $T_\mathrm{a}$, where $D$ denotes the fiber diameter (i.e., $A = \pi D^2 /4$). Furthermore, $g$ denotes the gravitational acceleration, $\gamma$ the surface tension and $G$ the melt shear modulus. In the amorphous constitutive equations  $\alpha$ is the Giesekus mobility parameter and $\zeta = N_0 l^2/3$ with $N_0$ number of flexible statistical links of length $l$ of one polymer chain. As for the semi-crystalline phase, the terms $\closureFz(S_{zz})$ and $\closureFr(S_{zz})$ are closure approximations arising from the underlying microstructural model, see Appendix~\ref{appendix_embedding}.
Moreover, $\sigma$ is the anisotropic drag coefficient, $F$ and $\delta$ model parameters for the semi-crystalline relaxation time, $\xi$ a parameter for the flow-enhanced crystallization and $\Phi_\infty$ the ultimate degree of crystallization. 

The two-phase model degenerates to a one-phase model for the amorphous melt if $S_{zz}=x\equiv 0$, $\lambda_\mathrm{sc} = 0$ are set. The one-phase model is a system of ordinary differential equations of first order, where the equation for the stored free energy $a$ decouples from the others.  The two-phase model, in contrast, contains a second derivative in $v_z$ entering via the stress tensor components. In a reformulation as first order system this implies the introduction of a new variable $\omega=\mathrm{d}{v_z}/\mathrm{d}z$.
In Doufas et al.\ \cite{doufas_simulation_2000-1} the one-phase model is coupled to the two-phase model as an interface problem where the interface corresponds to the point of onset of crystallization that is determined by a certain melt temperature. In Shrikhande et al.\ \cite{shrikhande_modified_2006} the two-phase model is considered along the whole spinline which implies the nozzle to act as point of onset of crystallization. 

The spinning setup provides indisputable boundary conditions at the nozzle for $A$, $v_z$, $T$ (and $a$) and at the outlet for $v_z$. At the point of onset of crystallization $x$ is zero. In \cite{doufas_simulation_2000-1} continuity of the unknowns appearing in the one-phase model is required at the interface. In addition, the acceleration $\omega$ is initialized with the momentum balance of the one-phase model and $S_{zz}$ with zero at the point of onset of crystallization. For the conformation tensor components, a viscous relation is prescribed at the nozzle. In \cite{shrikhande_modified_2006} the initialization of the acceleration $\omega$ is kept, whereas the other two conditions are replaced by expressions that are based on a force balance at the point of onset of crystallization. One of these expressions particularly contains a free parameter that has be chosen as suitable guess for the respective value of $\mathrm{d}{\omega}/\mathrm{d}z$.
So far, the posing of the last three boundary conditions has been rather heuristically motivated in all model variants available in literature and is under discussion. As the choice crucially affects the performance of the numerical solvers, we aim for boundary conditions that are mathematically consistent to the model class and prevent the occurrence of artificial layers.

\section{Asymptotic consideration of two-phase fiber model class} \label{sec_asymptotic}
The focus of this section is on the derivation of asymptotically justified boundary conditions and the formulation of consistent boundary value problems (BVP) for the two-phase fiber model class.

\begin{table}
\centering
\begin{tabular}{|l r@{\,}c@{\,}l l|}
\hline
\multicolumn{5}{|l|}{\textbf{Reference values}}							\\
Description					& \multicolumn{3}{l}{Formula}					& Unit	\\
\hline
Length						& $z_0$			&$=$& $L$						& m		\\
Cross-sectional area		& $A_0$			&$=$& $A_\text{in}=\pi D_\text{in}^2/4$			& $\tn{m}^2$ \\
Diameter					& $D_0$			&$=$& $\sqrt{A_0}$				& m		\\
Velocity					& $v_0$			&$=$& $v_\text{in}$					& m/s	\\
Temperature					& $T_0$			&$=$& $T_\text{in}$					& K		\\
Configuration				& $c_0$			&$=$& $\zeta$					& $\tn{m}^2$ \\
Free energy					& $a_0$			&$=$& $G$						& Pa	\\
Density						& $\rho_0$		&$=$& $\rho(T_\text{in},0)$			& kg/$\tn{m}^3$ \\
Viscosity					& $\mu_0$		&$=$& $\mu(T_\text{in})$				& Pa s 	\\
Heat capacity				& $C_{\mathrm{p},0}$		&$=$& $C_\mathrm{p}(T_\text{in},0)$			& J/(kg K) \\
Heat of fusion				& $\Delta H_{\mathrm{f},0}$ &$=$& $\Delta H_\mathrm{f}(T_\text{in},0)$	& J/kg	\\
Heat transfer coefficient	& $h_0$			&$=$& $h(v_\text{in},D_\text{in},T_\text{in})$		& W/($\tn{m}^2$ K) \\
Stress						& $\tau_0$		&$=$& $\mu_0 v_0 / z_0$			& Pa	\\
Relaxation time				& $\lambda_0$	&$=$& $\mu_0/G$				& s		\\
Crystallization rate		& $K_0$			&$=$& $K_\text{max}$					&1/s	\\
\hline
\multicolumn{1}{l}{\phantom{placeholder}} \\
\hline
\multicolumn{5}{|l|}{\textbf{Dimensionless numbers}}							\\
Description				& \multicolumn{3}{l}{Formula}							&\\
\hline
Slenderness				& $\epsilon$		&$=$& $ D_0/z_0$					&\\
Reynolds				& $\tn{Re}$			&$=$& $\rho_0 v_0 z_0/\mu_0$		&\\
Froude					&$\tn{Fr}$			&$=$& $v_0/\sqrt{z_0\,g}$			&\\
Capillary				&$\tn{Ca}$			&$=$& $\mu_0 v_0 / \gamma$		&\\
Stanton					&$\tn{St}$			&$=$& $h_0/(C_{\mathrm{p},0} \rho_0 v_0)$	&\\
Eckert					&$\tn{Ec}$			&$=$& $v_0^2/(C_{\mathrm{p},0}T_0)$	&\\
Jakob					&$\tn{Ja}$			&$=$& $C_{\mathrm{p},0}T_0/ (\Delta H_{\mathrm{f},0})$&\\
Deborah					&$\tn{De}$			&$=$& $\lambda_0v_0/z_0$				&\\
Damk\"ohler				&$\tn{Da}_\tn{I}$	&$=$& $K_0 z_0 / v_0$				&\\
Draw ratio					&$\tn{Dr}$			&$=$& $v_{\mathrm{out}}/v_\mathrm{in}$ 				&\\
\hline
\end{tabular}
\vspace{.3cm}
\caption{Reference values used for non-dimensionalization and resulting characteristic dimensionless numbers. The indices $_\text{in}$ and $_\text{out}$ indicate boundary data at inlet and outlet, respectively.}
\label{tab_nondim}
\end{table}

\subsection{Non-dimensionalization}
The two-phase fiber model class \eqref{eq:model} assumes a steady spinning-setup. Thus, the mass flux is constant, i.e., $W = (v_z \rho A)(z)=\text{const}$ for all $z\in [0,L]$, and the cross-sectional area $A$ can be expressed by the axial velocity $v_z$ and is no longer an unknown. Our strategy to make the fiber model dimensionless is essentially for the subsequent asymptotic consideration. In particular it differs from the one in \cite{doufas_simulation_2000-1, shrikhande_modified_2006}.  For each dimensional quantity $y$ we introduce a dimensionless one $y^*$ as $y^*(z^*) = y(z_0\,z^*)/y_0$ with reference value $y_0$. The respective reference values and the resulting dimensionless numbers are listed in Table \ref{tab_nondim}. Consequently, $z^* \in [0,1]$ and $W^*=1$ hold. To simplify the notation we suppress the index $^*$ from now on.
The dimensionless model system is then given by
\begin{subequations}\label{eq:dimensionless_2}
\begin{align}
\tn{Re} \; \deriv{v_z}	&= \ddz \bigg( \frac{\tau_{zz}-\tau_{rr}}{\rho v_z }\bigg) - \pi \frac{1}{\epsilon^2} B\mu_\mathrm{a} \left(v_z-v_\mathrm{a}^\parallel\right) + \frac{\tn{Re}}{\tn{Fr}^2} \frac{1}{v_z} + \sqrt{\pi} \frac{1}{\epsilon \tn{Ca}} \ddz \left(\frac{1}{\sqrt{\rho v_z }}\right), \label{eq_momentum_nondim} \\
\deriv{T}				&= - 2 \sqrt{\pi} \frac{\tn{St}}{\epsilon} \frac{h}{C_\mathrm{p} \sqrt{\rho v_z }} (T-T_\mathrm{a}) + \frac{\tn{Ec}}{\tn{Re}} \frac{\tau_{zz}-\tau_{rr}}{C_\mathrm{p} \rho v_z }\deriv{v_z} + \frac{1}{\tn{Ja}} \frac{\Phi_{\infty} \Delta H_\mathrm{f}}{C_\mathrm{p}} \deriv{x}, \\
\deriv{c_{zz}}			&= 2\frac{c_{zz}}{v_z} \deriv{v_z} - \frac{1}{\tn{De}} \frac{1}{\lambda v_z } \bigg(1-\alpha + \alpha \frac{c_{zz}}{1-x} \bigg) \bigg(\frac{c_{zz}}{(1-x)^2} - \frac{1}{1-x} \bigg), \\
\deriv{c_{rr}}			&= -\frac{c_{rr}}{v_z} \deriv{v_z} - \frac{1}{\tn{De}} \frac{1}{\lambda v_z } \bigg(1-\alpha + \alpha \frac{c_{rr}}{1-x} \bigg) \bigg(\frac{c_{rr}}{(1-x)^2} - \frac{1}{1-x} \bigg), \\
\delta \; \deriv{S_{zz}}		&= - \frac{\sigma}{\tn{De}} \frac{S_{zz}}{ \lambda \exp(Fx)v_z} + \delta \frac{1}{v_z} \left(2 S_{zz} + \frac{2}{3} - 2 \closureFz(S_{zz})\right)\deriv{v_z}, \label{eq_orientational_nondim}\\
\deriv{x} 				&= \tn{Da}_{\tn{I}}K \,\frac{1-x}{v_z}, \\
\deriv{a}				&= - \frac{1}{\tn{De}}\frac{a}{ \lambda v_z (1-x)^2} + \frac{c_{zz} - c_{rr}}{v_z(1-x)} \deriv{v_z}, \label{eq_free_energy_nondim}
\end{align}
with
\begin{align}
\tau_{zz}	= \frac{1}{\tn{De}} \left(\frac{c_{zz}}{1-x} -1 + 3 S_{zz}\right) + 6 \delta \lambda \exp(Fx) \,\closureFz(S_{zz}) \deriv{v_z}, \label{eq_dtau_nondim_1} \\
\tau_{rr}	= \frac{1}{\tn{De}} \left(\frac{c_{rr}}{1-x} -1 - \frac{3}{2} S_{zz}\right) + 6 \delta \lambda \exp(Fx) \,\closureFr(S_{zz}) \deriv{v_z}. \label{eq_dtau_nondim_2}
\end{align}
\end{subequations}

The system contains a second derivative of $v_z$, when inserting the equations for the stress tensor components  $\tau_{zz}$ and $\tau_{rr}$ into the momentum equation \eqref{eq_momentum_nondim}. We reformulate it as an explicit first order system by introducing the new independent variable, the acceleration, $\omega = \mathrm{d}v_z /\mathrm{d}z$. Equation~\eqref{eq_momentum_nondim} then becomes 
\begin{subequations} \label{eq:system1}
\begin{align}
\deriv{v_z}		&= \omega, \\
\begin{split} \label{eq_dgl_omega}
\delta \,\abk(T, S_{zz}, x) \deriv{\omega}	&= \tn{Re} \; \rho v_z  \omega + \pi \frac{1}{\epsilon^2} B \mu_\mathrm{a} \rho v_z  \left(v_z-v_\mathrm{a}^\parallel\right) - \frac{\tn{Re}}{\tn{Fr}^2} \rho \\
						& \quad + \left(\frac{\sqrt{\pi}}{2} \frac{1}{\epsilon \tn{Ca}} \frac{1}{\sqrt{ \rho v_z}} \right) \left( \rho \omega+ \deriv{\rho} v_z \right) \\
						& \quad + \left(\frac{1}{\tn{De}}\left(\frac{c_{zz}-c_{rr}}{1-x} + \frac{9}{2} S_{zz} \right) + \delta \omega  \abk(T, S_{zz}, x)  \right) \left(\frac{\omega}{v_z} + \deriv{\rho} \frac{1}{\rho} \right) \\
						& \quad - \frac{1}{\tn{De}} \left(\ddz \left(\frac{c_{zz}-c_{rr}}{1-x}\right) + \frac{9}{2} \deriv{S_{zz}} \right) - \delta \omega \ddz \abk(T, S_{zz}, x)
\end{split}
\end{align}
\end{subequations}
with $\abk(T, S_{zz}, x) = 6 \lambda(T) \exp(Fx) \left(\closureFz(S_{zz})-\closureFr(S_{zz})\right)$.

\subsection{Asymptotically justified boundary conditions}
The dimensionless first order differential system contains a small model parameter $\delta$, $0<\delta\ll 1$, we make use of in our asymptotic consideration. The parameter enters via the semi-crystalline relaxation time. According to \cite{doufas_continuum_1998, doufas_simulation_2000-1}, it is chosen very small to achieve small relaxation times for the infinitesimal semi-crystalline phase at the beginning.  The parameter occurs explicitly in the equations  \eqref{eq_orientational_nondim} and \eqref{eq_dgl_omega} for the orientation tensor component $S_{zz}$ and the acceleration $\omega$, respectively. We expand $S_{zz}$  in a regular power series with respect to $\delta$, i.e., $S_{zz}=S_{zz}^{(0)}+\delta\, S_{zz}^{(1)}+\mathcal{O}(\delta^2)$. Inserting this expansion into \eqref{eq_orientational_nondim}, we find $S_{zz}^{(0)}=0$ in leading order. Hence, we introduce the scaled variable $S=S_{zz}/\delta$. We collect the variables as 
\begin{align*}
\bs{x} = \left(v_z, T, c_{zz}, c_{rr}, x, a \right), \qquad \qquad \bs{y} = (S,\omega)
\end{align*}
and rearrange the equation system \eqref{eq:dimensionless_2} under consideration of \eqref{eq:system1} in the form
\begin{subequations}
\begin{align}
\deriv{\bs{x}} &= \bs{F}(\bs{x},\bs{y};\delta), \label{eq_ode_dynamic}\\
\delta \, \begin{pmatrix}1& 0 \\ 0&\abk(\bs{x},\bs{y};\delta)\end{pmatrix} \cdot \deriv{\bs{y}} &= \bs{G}(\bs{x},\bs{y};\delta) \label{eq_ode_algebraic}
\end{align}
\end{subequations}
with $\abk(\bs{x},\bs{y};\delta) = 6 \lambda(T) \exp(Fx) \left(\closureFz(\delta S) - \closureFr(\delta S\right))$.
In the asymptotic limit $\delta \rightarrow 0$, the fiber model degenerates to a differential-algebraic system of index 1, where the equations \eqref{eq_ode_algebraic} for $\bs y=(S,\omega)$ become algebraic relations, i.e., 
\begin{align}\label{eq:constraint}
\bs{0} = \bs{G}(\bs{x},\bs{y};0). 
\end{align}
So, $\bs{x}$ are the dynamic variables and $\bs{y}$ the algebraic ones.

In the context of perturbation theory, the fiber model is described by a $\delta$-perturbed differential system of first order. Boundary layers or even discontinuities in the solutions are to be expected in the case of a singular perturbation, when the posed boundary conditions are not consistent to the algebraic relations \eqref{eq:constraint}. This can be observed in, e.g., \cite{doufas_simulation_2000-1}, where a vanishing orientation tensor component is prescribed. We propose a regular perturbation. To set up suitable boundary conditions we use the asymptotic derivation presented in Appendix~\ref{appendix_asymptotic}. Imposing the algebraic relations \eqref{eq:constraint} as boundary conditions yields a regular perturbation of zeroth order. We impose the $\delta$-extended relations 
\begin{align} \label{eq:first_order_bc}
\bs{0} = \delta \, \partial_{\bs{x}} \bs{G}(\bs{x},\bs{y};0) \cdot \bs{F}(\bs{x},\bs{y};\delta) + \partial_{\bs{y}} \bs{G}(\bs{x},\bs{y};0) \cdot \begin{pmatrix}1& 0 \\ 0&1/\abk(\bs{x},\bs{y};\delta)\end{pmatrix} \cdot \bs{G}(\bs{x},\bs{y};\delta)
\end{align}
as boundary conditions which implies a regular perturbation of first order in $\delta$ (i.e., consistency up to terms of $\mathcal{O}(\delta^2$)) and results in regular solutions without any layers as we will show.

\begin{remark} \label{rem:1}
The function $\bs{G}$ in \eqref{eq_ode_algebraic} has a special structure in $\delta$, i.e.,
\begin{align} \label{eq:S_w}
\bs{G}(\bs{x},\bs{y};\delta) =
\begin{pmatrix}
\mathcal{S}_0(\bs{x},\bs{y}) + \delta \ \mathcal{S}_1(\bs{x},\bs{y};\delta) \\
\Omega_0(\bs{x},\bs{y}) + \delta \ \Omega_1(\bs{x},\bs{y};\delta)
\end{pmatrix}
\end{align}
with
\begin{alignat*}{3}
&\mathcal{S}_0(\bs{x},\bs{y}) &&= - \frac{\sigma}{\tn{De}} \frac{S}{\lambda \exp(Fx)v_z } + \frac{2}{5} \frac{\omega}{v_z}, \\
&\mathcal{S}_1(\bs{x},\bs{y};\delta) &&= 2 (S - \closureFzTilde(S;\delta)) \frac{\omega}{v_z}, \quad  \quad \closureFzTilde(S;\delta)			= -\frac{81}{8}\delta^4 S^5 + \frac{675}{56}\delta^3 S^4 - \frac{36}{35}\delta^2 S^3 - \frac{9}{10}\delta S^2 + \frac{11}{14}S, \\
&\Omega_0(\bs{x},\bs{y}) &&= \tn{Re} \; \rho v_z  \omega + \pi \frac{1}{\varepsilon^2} B\mu_\mathrm{a} \rho  v_z \left(v_z-v_\mathrm{a}^\parallel\right) - \frac{\tn{Re}}{\tn{Fr}^2} \rho + \left(\frac{\sqrt{\pi}}{2} \frac{1}{\varepsilon \tn{Ca}} \frac{1}{\sqrt{\rho v_z }}\right) \left(\rho \omega  + \deriv{\rho} v_z \right) \\
						&&& \quad + \frac{1}{\tn{De}}\left(\frac{c_{zz}-c_{rr}}{1-x}\right)\left(\frac{\omega}{v_z} + \deriv{\rho} \frac{1}{\rho} \right) - \frac{1}{\tn{De}} \ddz \left(\frac{c_{zz}-c_{rr}}{1-x}\right) - \frac{9}{2}\frac{1} {\tn{De}} \mathcal{S}_0(\bs{x},\bs{y}), \\
&\Omega_1(\bs{x},\bs{y};\delta) &&= \left(\frac{9}{2}\frac{1}{\tn{De}} S + \abk(\bs{x},\bs{y};\delta) \omega\right) \left(\frac{\omega}{v_z} + \deriv{\rho} \frac{1}{\rho} \right) - \frac{9}{2}\frac{1}{ \tn{De}} \mathcal{S}_1(\bs{x},\bs{y};\delta) - \deriv{\abk}(\bs{x},\bs{y};\delta) \,\omega.
\end{alignat*}
Note that in $\Omega_0$ and $\Omega_1$ the notation of the derivative is used as abbreviation for the respective expression in terms of the variables $(\bs{x},\bs{y})$.

In the asymptotic limit ($\delta=0$) the algebraic relations particularly read \begin{align}\label{eq:algebraic}
\bs{0} = \bs{G}(\bs{x},\bs{y};0) =
(\mathcal{S}_0(\bs{x},\bs{y}),
\Omega_0(\bs{x},\bs{y}))^T.  
\end{align}
The asymptotic consideration reveals that for small semi-crystalline relaxation times (for small $\delta$) the orientation tensor component is proportional to the acceleration which mainly arises from the amorphous melt,
\begin{align*}
S_{zz}=\delta S= \delta \frac{2}{5} \frac{\mathrm{De}}{\sigma} \lambda \exp(Fx)\,\omega +\mathcal{O}(\delta^2), \qquad \omega=\omega^{(0)}+\mathcal{O}(\delta).
\end{align*}
Here, $\omega^{(0)}$ denotes the acceleration associated to $\Omega_0(\bs{x},\bs{y})=0$.
\end{remark}

\subsection{Regularly perturbed boundary value problems} 
For the flow-enhanced crystallization in fiber spinning we adjust the model variants of Doufas et al.\ \cite{doufas_simulation_2000-1} and Shrikhande et al.\  \cite{shrikhande_modified_2006}. We present regularly perturbed boundary value problems and discuss their differences and advantages over the models from the original literature.

\subsubsection*{Non-crystallizing fiber}
A non-crystallizing fiber model results from the two-phase model \eqref{eq:dimensionless_2} in the limit for vanishing semi-crystalline relaxation time and Damk\"ohler number, i.e., $\delta= 0$ and $\tn{Da}_\tn{I} = 0$. In the limit the algebraic relations \eqref{eq:algebraic} and $\mathrm{d}x/\mathrm{d}z \equiv 0$ hold, moreover the equation for the stored free energy $a$ decouples from the system. The crystallinity $x$ and the orientation tensor component $S_{zz}$ are zero along the spinline. The remaining first order differential system for $v_z$, $T$, $c_{zz}$ and $c_{rr}$ embedded in a boundary value problem for fiber spinning reads as follows.
\begin{system}[BVP for non-crystallizing fiber] \label{system_reduced_model}
\begin{align*}
\deriv{v_z} 	&= \omega_\mathrm{am},\\
\deriv{T}		&= - 2 \sqrt{\pi} \frac{\tn{St}}{\varepsilon} \frac{h}{C_\mathrm{p} \sqrt{ \rho v_z}} (T-T_\mathrm{a}) + \frac{\tn{Ec}}{\tn{Re De}} \frac{c_{zz}-c_{rr}}{C_\mathrm{p} \rho  v_z}\omega_\mathrm{am}, \\
\deriv{c_{zz}}	&= 2 \frac{c_{zz}}{v_z}  \omega_\mathrm{am} - \frac{1}{\tn{De}} \frac{1}{ \lambda v_z} (1-\alpha + \alpha c_{zz}) (c_{zz} - 1), \\
\deriv{c_{rr}}	&= - \frac{c_{rr}}{v_z}  \omega_\mathrm{am} - \frac{1}{\tn{De}} \frac{1}{\lambda v_z } (1-\alpha + \alpha c_{rr} ) (c_{rr} - 1).
\end{align*}
Boundary conditions at inlet $z=0$ and outlet $z=1$:
\begin{equation*}
v_z(0) = 1, \qquad v_z(1) = \tn{Dr}, \qquad T(0) = 1, \qquad c_{zz}(0) + 2c_{rr}(0) = 3.
\end{equation*}
\end{system}
We assume here the resolvability of the algebraic relation $ \Omega_0(\bs{x},\bs{y})=0$ for $\omega$ (cf.\ Remark~\ref{rem:1}). With $\mathcal{S}_0(\bs{x},\bs{y})=0$, the expression $\omega_\mathrm{am}$ stands for the acceleration arising in the amorphous melt, i.e.,
\begin{align}\label{eq:omega_am}\nonumber
\omega_\mathrm{am}&= \Bigg( \tn{Re} \; \rho v_z - \frac{1}{\tn{De}} \frac{c_{zz}+2c_{rr}}{v_z} + \frac{\sqrt{\pi}}{2} \frac{1}{\varepsilon \tn{Ca}} \sqrt{\frac{\rho}{v_z}} +\frac{\tn{d}\rho}{\tn{d}T} \; \mathcal{V}_1 \Bigg)^{-1} \\
& \quad \; \Bigg(- \pi \frac{1}{\varepsilon^2} B \mu_\mathrm{a}\rho  v_z \left(v_z - v_\mathrm{a}^\parallel\right) + \frac{\tn{Re}}{\tn{Fr}^2} \rho + \frac{\tn{d}\rho}{\tn{d}T} \; \mathcal{V}_2 \\\nonumber
& \qquad + \frac{1}{\tn{De}^2}\frac{1}{ \lambda v_z} \bigg(\left(1-\alpha+\alpha c_{rr}\right)\left(c_{rr}-1\right) - \left(1-\alpha+\alpha c_{zz}\right)\left(c_{zz}-1\right)\bigg) \Bigg), \\\nonumber
\mathcal{V}_1 &= \frac{\tn{Ec}}{\tn{Re\,De}} \frac{c_{zz}-c_{rr}}{C_\mathrm{p} \rho \sqrt{\rho v_z}} \left(\frac{\sqrt{\pi}}{2} \frac{1}{\varepsilon \tn{Ca}}  + \frac{1}{\tn{De}} \frac{c_{zz}-c_{rr}}{\sqrt{\rho v_z}}\right), \\\nonumber
\mathcal{V}_2 &= \sqrt{\pi} \frac{\tn{St}}{\varepsilon} \frac{h}{C_\mathrm{p} \rho} \left(T-T_\mathrm{a}\right) \left(\sqrt{\pi}\frac{1}{\varepsilon \tn{Ca}} + \frac{2}{\tn{De}} \frac{c_{zz}-c_{rr}}{\sqrt{\rho v_z }}\right).
\end{align}
The boundary value problem is obviously unperturbed. The boundary conditions reflect the spinning set-up with prescribed velocities at inlet and outlet (with draw ratio $\mathrm{Dr}$). The stated relation for the conformation tensor components accounts for the viscous fiber behavior at the inlet, cf.\  \cite{doufas_simulation_2000-1}. System~\ref{system_reduced_model} with constant fiber density $\rho$ particularly coincides with the one-phase model that is used in \cite{doufas_simulation_2000-1} before the onset of crystallization. The consideration of a temperature-dependent fiber density results here in the additional expressions $\mathcal{V}_1$ and $\mathcal{V}_2$.

\subsubsection*{Flow-enhanced stress-driven crystallization with variable onset along the spinline}
For the fiber spinning model with stress-driven crystallization with variable onset, we propose two regularly perturbed boundary value problems with a free interface. The interface is the point of crystallization onset $z^\lozenge\in [0,1]$ that is an unknown of the problem and implicitly determined by a certain reached melt temperature, i.e., $T(z^\lozenge)=T^\lozenge$. The further variables are $v_z$, $T$, $c_{zz}$ and $c_{rr}$ before as well as after onset of crystallization, and additionally $x$, $S$ and $\omega$ after onset of crystallization.
\begin{system}[Free interface-BVP for two-phase flow with variable onset of crystallization] \label{system_final_model_doufas} \quad \\
Before onset of crystallization, $z \in [0,z^\lozenge)$:
\begin{align*}
\deriv{v_z} 	&= \omega_\mathrm{am}, \\
\deriv{T}		&= - 2 \sqrt{\pi} \frac{\tn{St}}{\varepsilon} \frac{h}{C_\mathrm{p} \sqrt{\rho v_z }} (T-T_\mathrm{a}) + \frac{\tn{Ec}}{\tn{Re De}} \frac{c_{zz}-c_{rr}}{C_\mathrm{p} \rho v_z}\omega_\mathrm{am}, \\
\deriv{c_{zz}}	&= 2 \frac{c_{zz}}{v_z} \omega_\mathrm{am} - \frac{1}{\tn{De}} \frac{1}{\lambda v_z} (1-\alpha + \alpha c_{zz}) (c_{zz} - 1 ), \\
\deriv{c_{rr}}	&= - \frac{c_{rr}}{v_z}\omega_\mathrm{am} - \frac{1}{\tn{De}} \frac{1}{\lambda v_z } (1-\alpha+ \alpha c_{rr}) (c_{rr} - 1).
\end{align*}
After onset of crystallization, $z \in (z^\lozenge, 1]$:
\begin{align*}
\deriv{v_z}									&= \omega, \\
\delta \,\abk(\bs{x},\bs{y};\delta) \deriv{\omega} 	&= \Omega_0(\bs{x},\bs{y}) + \delta \; \Omega_1(\bs{x},\bs{y};\delta), \\
\deriv{T}				&= - 2 \sqrt{\pi} \frac{\tn{St}}{\epsilon} \frac{h}{C_\mathrm{p} \sqrt{\rho v_z }} (T-T_\mathrm{a}) + \frac{\tn{Ec}}{\tn{Re}} \frac{\tau_{zz}-\tau_{rr}}{C_\mathrm{p} \rho v_z}\omega + \frac{\tn{Da}_{\tn{I}}}{\tn{Ja}} \frac{\Phi_{\infty} \Delta H_\mathrm{f}K}{C_\mathrm{p}} \,\frac{1-x}{v_z}, \\
\deriv{c_{zz}}			&= 2\frac{c_{zz}}{v_z}\omega - \frac{1}{\tn{De}} \frac{1}{\lambda v_z} \bigg(1-\alpha + \alpha \frac{c_{zz}}{1-x} \bigg) \bigg(\frac{c_{zz}}{(1-x)^2} - \frac{1}{1-x} \bigg), \\
\deriv{c_{rr}}			&= -\frac{c_{rr}}{v_z}\omega - \frac{1}{\tn{De}} \frac{1}{\lambda v_z} \bigg(1-\alpha + \alpha \frac{c_{rr}}{1-x} \bigg) \bigg(\frac{c_{rr}}{(1-x)^2} - \frac{1}{1-x} \bigg), \\
\delta \, \deriv{S} 							&= \mathcal{S}_0(\bs{x},\bs{y}) + \delta \; \mathcal{S}_1(\bs{x},\bs{y};\delta), \\
\deriv{x} 				&= \tn{Da}_{\tn{I}}\, K(T,\tau_{zz},\tau_{rr})\,\frac{1-x}{v_z}
\end{align*}
with $\omega_\mathrm{am}$ of \eqref{eq:omega_am}, the asymptotic $\delta$-associated expressions of \eqref{eq:S_w} and 
\begin{align*}
\tau_{zz}	&= \frac{1}{\tn{De}} \left(\frac{c_{zz}}{1-x} -1 + 3 \delta S\right) + 6 \delta \lambda \exp(Fx) \,\closureFz(\delta S) \, \omega, \\
\tau_{rr}	&= \frac{1}{\tn{De}} \left(\frac{c_{rr}}{1-x} -1 - \frac{3}{2} \delta S\right) + 6 \delta \lambda \exp(Fx) \,\closureFr(\delta S)  \,\omega. 
\end{align*}
Interface condition for onset of crystallization: $T(z^\lozenge)= T^\lozenge$.\\
Boundary conditions at inlet $z=0$, outlet $z=1$ and interface $z=z^\lozenge$:
\begin{align*}
v_z(0)&= 1,	 \qquad \qquad v_z(1)= \tn{Dr},	\qquad \qquad   T(0)= 1,	\qquad  \qquad c_{zz}(0) + 2c_{rr}(0)	= 3, \\
x(z^\lozenge)&=0, \qquad \qquad  \hspace*{3.2cm} \lim_{z \uparrow z^\lozenge} (v_z, T, c_{zz}, c_{rr}) (z)= \lim_{z \downarrow z^\lozenge} (v_z, T, c_{zz}, c_{rr}) (z),
\end{align*}
\underline{Case A:} \hspace*{1.cm} $
\omega(z^\lozenge)	= \lim_{z \uparrow z^\lozenge} \deriv{v_z}(z)$,  \hspace*{1.75cm} $ \mathcal{S}_0(\bs{x}(z^\lozenge),\bs{y}(z^\lozenge)) = 0$,\\
\underline{Case B:} \\
\hspace*{0.5cm} $
\bs{0} = \delta \, \partial_{\bs{x}} \bs{G}(\bs{x},\bs{y};0) \cdot \bs{F}(\bs{x},\bs{y};\delta) +\partial_{\bs{y}} \bs{G}(\bs{x},\bs{y};0) \cdot \begin{pmatrix}1& 0 \\ 0&1/\abk(\bs{x},\bs{y};\delta)\end{pmatrix} \cdot \bs{G}(\bs{x},\bs{y};\delta)\bigg|_{(\bs{x},\bs{y})=(\bs{x}(z^\lozenge),\bs{y}(z^\lozenge))}.
$
\end{system}
The boundary conditions at inlet and outlet reflect the spinning set-up. The conditions at the interface ensure the continuity of the quantities that occur along the whole spinline and the consistent initialization of the crystallinity.
Note that as for the asymptotically justified conditions at the interface we distinguish two cases. Case A represents zeroth-order boundary conditions that coincide with the algebraic relations \eqref{eq:constraint}. Acceleration and orientation stress tensor are initialized with respect to the state before onset of crystallization where $\delta=0$ and $\tn{Da}_\tn{I}=0$.
From $\Omega_0(\bs{x}(z^\lozenge),\bs{y}(z^\lozenge)) = 0$ and $\mathcal{S}_0(\bs{x}(z^\lozenge),\bs{y}(z^\lozenge)) = 0$ with $\tn{Da}_\tn{I}=0$ and $x(z^\lozenge)=0$, we conclude that 
\begin{align*}
\omega(z^\lozenge)=\omega_\mathrm{am}(z^\lozenge), \qquad  \qquad S_{zz}(z^\lozenge)= \delta  \, \frac{2}{5} \,\frac{\mathrm{De}}{\sigma} \,\lambda(T^\lozenge)\, \omega_\mathrm{am}(z^\lozenge).
\end{align*}
Thus, acceleration and orientation tensor component at the point of crystallization onset are determined by the acceleration of the amorphous melt (cf.\ \eqref{eq:omega_am} and Remark~\ref{rem:1}).
As continuity of the acceleration is imposed, continuous derivatives for all quantities that occur along the whole spinline can be expected. However, this might go with a moderate layering in the derivatives of the algebraic variables, since terms of the order $\mathcal{O}(\delta)$ are ignored in the conditions.
Case~B represents first-order boundary conditions where higher order terms are taken into account by imposing \eqref{eq:first_order_bc}. We expect no layering in the derivatives of the algebraic variables, while
discontinuities in the derivatives of the other quantities can occur at the interface. These jumps are (mathematically) reasonable due to the switching of the model equations at the interface (from $\delta=\tn{Da}_\tn{I}=0$ to $\delta\neq 0$, $\tn{Da}_\tn{I}\neq0 $), they are common in interface problems in literature. We point out that both cases are viable possibilities to close the interface-boundary value problem, both imply regularly perturbed systems. The resulting differences in the solutions are investigated in Section~\ref{sec:numerics}.

In Doufas et al.\ \cite{doufas_simulation_2001, doufas_simulation_2000-1, doufas_simulation_2000} the flow-enhanced crystallization is handled via a switching approach in a numerical shooting method. At first, the one-phase model is used until $z^\lozenge$ is found via $T(z^\lozenge)=T^\lozenge$. Then, the two-phase model with stress-driven crystallization is applied to the spinline part $[z^\lozenge,1]$, using $\omega(z^\lozenge)=\omega_\mathrm{am}(z^\lozenge)$ and $S_{zz}(z^\lozenge)=0$. This procedure involves inaccuracies regarding the position of onset of crystallization, since effects of the two-phase flow do not enter equally in the determination of $z^\lozenge$. Moreover, the condition $S_{zz}(z^\lozenge)=0$ yields layering due to the present small $\delta$. 

\begin{remark} \label{rem:2}
Due to the stress-dependent crystallization rate $K(T,\tau_{zz},\tau_{rr})$, the crystallinity $x$ can approach one in certain spinning setups. The raising singularities in the right hand side of the model equations cause then the break-down of the numerical solvers. Whereas Doufas et al.\ performed a model switching by setting $K \equiv 0$ for $x \geq x_\text{crit}$ and accepted a non-differentiable solution at the switching point, we propose a regularization to ensure smooth solutions and, consequently, good solver convergence behavior. Moreover, it is beneficial for the numerical treatment to transform System~\ref{system_final_model_doufas} such that the free interface becomes a fixed boundary and the unknown $z^\lozenge$ is shifted into the differential equations. See Appendix~\ref{subsec:System2} for details.
\end{remark}

\subsubsection*{Flow-enhanced crystallization driven by free stored energy with onset at inlet}
For the spinning of the two-phase fiber flow with free stored energy-driven crystallization with onset at the inlet, we propose the following regularly perturbed boundary value problem for $v_z$, $T$, $c_{zz}$, $c_{rr}$, $x$, $S$, $\omega$ and the free stored energy $a$.
\begin{system}[BVP for two-phase flow with onset of crystallization at inlet] \label{system_final_model_shrikhande}
\begin{align*}
\deriv{v_z}									&= \omega, \\
\delta \, \abk(\bs{x},\bs{y};\delta) \deriv{\omega} 	&= \Omega_0(\bs{x},\bs{y}) + \delta \; \Omega_1(\bs{x},\bs{y};\delta), \\
\deriv{T}				&= - 2 \sqrt{\pi} \frac{\tn{St}}{\epsilon} \frac{h}{C_\mathrm{p} \sqrt{\rho v_z }} (T-T_\mathrm{a}) + \frac{\tn{Ec}}{\tn{Re}} \frac{\tau_{zz}-\tau_{rr}}{C_\mathrm{p}\rho v_z } \omega + \frac{\tn{Da}_{\tn{I}}}{\tn{Ja}} \frac{\Phi_{\infty} \Delta H_\mathrm{f}K}{C_\mathrm{p}} \,\frac{1-x}{v_z}, \\
\deriv{c_{zz}}			&= 2\frac{c_{zz}}{v_z} \omega- \frac{1}{\tn{De}} \frac{1}{ \lambda v_z}  \bigg(1-\alpha + \alpha \frac{c_{zz}}{1-x}  \bigg) \bigg(\frac{c_{zz}}{(1-x)^2} - \frac{1}{1-x} \bigg), \\
\deriv{c_{rr}}			&= -\frac{c_{rr}}{v_z}  \omega - \frac{1}{\tn{De}} \frac{1}{\lambda v_z } \bigg(1-\alpha + \alpha \frac{c_{rr}}{1-x} \bigg) \bigg(\frac{c_{rr}}{(1-x)^2} - \frac{1}{1-x} \bigg), \\
\delta \,\deriv{S} 							&= \mathcal{S}_0(\bs{x},\bs{y}) + \delta \; \mathcal{S}_1(\bs{x},\bs{y};\delta), \\
\deriv{x} 				&= \tn{Da}_{\tn{I}}\,K(T,a)\,\frac{1-x}{v_z}, \\
\deriv{a}				&= - \frac{1}{\tn{De}}\frac{a}{\lambda v_z  (1-x)^2} + \frac{c_{zz} - c_{rr}}{v_z(1-x)}  \omega
\end{align*}
with the asymptotic $\delta$-associated expressions of \eqref{eq:S_w} and 
\begin{align*}
\tau_{zz}	&= \frac{1}{\tn{De}} \left(\frac{c_{zz}}{1-x} -1 + 3 \delta S\right) + 6 \delta \lambda \exp(Fx) \,\closureFz(\delta S)\,  \omega, \\
\tau_{rr}	&= \frac{1}{\tn{De}} \left(\frac{c_{rr}}{1-x} -1 - \frac{3}{2} \delta S\right) + 6 \delta \lambda \exp(Fx) \,\closureFr(\delta S) \, \omega. 
\end{align*}
Boundary conditions at inlet $z=0$ and outlet $z=1$:
\begin{alignat*}{7}
v_z(0)	&= 1,	& \qquad v_z(1)	&= \tn{Dr},	& \qquad T(0)	&= 1, \\
x(0) 	&= 0,		& \qquad a(0)	&= 0, & \qquad c_{zz}(0) + 2c_{rr}(0)	&= 3,
\end{alignat*}
\hspace*{0.5cm} $\bs{0} = \delta \, \partial_{\bs{x}} \bs{G}(\bs{x},\bs{y};0) \cdot \bs{F}(\bs{x},\bs{y};\delta) +\partial_{\bs{y}} \bs{G}(\bs{x},\bs{y};0) \cdot \begin{pmatrix}1& 0 \\ 0&1/\abk(\bs{x},\bs{y};\delta)\end{pmatrix} \cdot \bs{G}(\bs{x},\bs{y};\delta)\bigg|_{(\bs{x},\bs{y})=(\bs{x}(0),\bs{y}(0))}$.
\end{system}
The spinning setup provides the boundary conditions for $v_z$, $T$ and $a$. As the point of crystallization onset is assumed to be in the nozzle, the crystallinity $x$ is zero here. The viscous relation for the conformation tensor components accounts for the microstructural behavior before spinning as in System~\ref{system_reduced_model} and System~\ref{system_final_model_doufas}. The two asymptotically justified boundary conditions of first order ensure a regular solution behavior for small $\delta$. 

In Shrikhande et al.\ \cite{shrikhande_modified_2006} the following boundary conditions are imposed in place of our three last described ones, i.e.,
\begin{align}\label{eq:bc_S}
\mathcal{C}_i(\bs{x}(0),\bs{y}(0);P^\star)=0, \qquad i=1,2,3
\end{align}
with the parameter $P^\star$ chosen as a guess for the unknown quantity $\mathrm{d}\omega/\mathrm{d}z\,(0)$ and 
\begin{align*}
	\mathcal{C}_1(\bs{x},\bs{y})\, \quad &= \tn{Re} \; \rho v_z  \omega - \frac{\tn{Re}}{\tn{Fr}^2} \rho  + \frac{\sqrt{\pi}}{2} \frac{1}{\epsilon \tn{Ca}} \sqrt{\frac{\rho}{v_z}}  \omega  + \frac{1}{\tn{De}}\left((c_{zz}-c_{rr}) \frac{\omega}{v_z}  - \ddz (c_{zz}-c_{rr}) \right),\\
	\mathcal{C}_2(\bs{x},\bs{y};P)&= \omega\left(\frac{\omega}{v_z}-\frac{1}{\lambda}\deriv{\lambda}\right)-P,\\
\mathcal{C}_3(\bs{x},\bs{y})\, \quad &= \frac{9}{2} \frac{1}{\tn{De}} \left( S_{zz} \frac{\omega}{v_z}-   \deriv{S_{zz}} \right)  -6\delta \lambda \omega \, \ddz{(\exp(Fx)\, (\mathcal{U}_z(S_{zz})-\mathcal{U}_r(S_{zz})))}. 
	\end{align*}
Note that the fiber density $\rho$ is taken as constant in \cite{shrikhande_modified_2006}.
The boundary conditions are motivated from a partitioning of \eqref{eq_dgl_omega} under some simplifying assumptions on the flow at the inlet.  We point out that the two-phase fiber model is not closed with \eqref{eq:bc_S} due to the additionally introduced parameter $P$. Its choice crucially affects the solution behavior as we will show. Moreover, when considering \eqref{eq:bc_S} with fixed $P^\star$ as an equation system for the unknown inlet boundary values $(c_{rr}, \omega, S_{zz})(0)$, e.g., in the context of a numerical shooting method as in \cite{shrikhande_modified_2006}, the nonlinear fully coupled system allows for several solutions. 

\begin{remark}[Shrikhande-like boundary conditions] \label{rem:shriklike_bc}
As for \eqref{eq:bc_S}, the mentioned problems can be also overcome when closing the model system by help of an additional equation for the free parameter $P$. Including the viscous relation for the conformation tensor components at the inlet, $\mathcal{C}_2(\bs{x}(0),\bs{y}(0);P)=0$ becomes an equation for $P$ and can be eliminated. Since
\begin{align*}
\mathcal{C}_1(\bs{x},\bs{y})&=\left[\Omega_0(\bs{x},\bs{y})+\frac{9}{2}\frac{1}{\tn{De}}\mathcal{S}_0(\bs{x},\bs{y})\right]_{\rho=\mathrm{const}, B=\tn{Da}_\tn{I}=0}=\omega_\mathrm{am}\,\bigg|_{\rho=\mathrm{const},B=0},\\
\mathcal{C}_3(\bs{x},\bs{y})&=\left[-\frac{9}{2}\frac{1}{\tn{De}}\mathcal{S}_0(\bs{x},\bs{y})+\delta \left(\Omega_1(\bs{x},\bs{y};\delta)-\abk{(\bs{x},\bs{y};\delta)}\, \omega \left(\frac{\omega}{v_z}-\frac{1}{\lambda}\deriv{\lambda}\right)\right)\right]_{\rho=\mathrm{const}}
\end{align*}
(cf.\ \eqref{eq:S_w}), the resulting equation system partially decouples and yields a unique expression for $(c_{rr}, \omega, S_{zz})(0)$. Rejecting the simplifications $\rho=\mathrm{const}$, $B=0$ motivates the following Shrikhande-like boundary conditions, i.e.,
\begin{align}\label{eq:bc_alternative}\nonumber
 c_{zz}(0) + 2c_{rr}(0)	= 3, \qquad \qquad \qquad \quad 
 \left[\Omega_0+\frac{9}{2}\frac{1}{\tn{De}} \mathcal{S}_0 \right ]_{\tn{Da}_\tn{I}=0}  (\bs{x}(0),\bs{y}(0)) &= 0,\\ 
\left[ -\frac{9}{2}\frac{1}{\tn{De}}\mathcal{S}_0+\delta \left(\Omega_1-\abk\omega \left(\frac{\omega}{v_z}-\frac{1}{\lambda}\deriv{\lambda}\right)\right)\right ]_{(\bs{x},\bs{y})=(\bs{x}(0),\bs{y}(0))}&=0
\end{align}
where the acceleration is initialized with respect to the amorphous melt, i.e., $\omega_\mathrm{am}(0)=0$, cf.\ \eqref{eq:omega_am}. Since the last condition contains additional terms of $\mathcal{O}(\delta)$, more appropriate approximations of $\mathrm{d}\omega/\mathrm{d}z$ near the inlet than with our zeroth-order boundary conditions (i.e., $\mathcal{S}_0(\bs{x}(0),\bs{y}(0)) = 0$ and $\Omega_0(\bs{x}(0),\bs{y}(0)) = 0$) can be expected for small  $\delta$. However, the approximation quality of our first-order boundary conditions cannot be reached due to the incomplete asymptotics (see Section~\ref{sec:numerics}). 
\end{remark}

\section{Impact on numerical solvers and results}\label{sec:numerics}

\begin{figure}[tbh]
\centering
\subfigure{\includegraphics[width=0.3\textwidth]{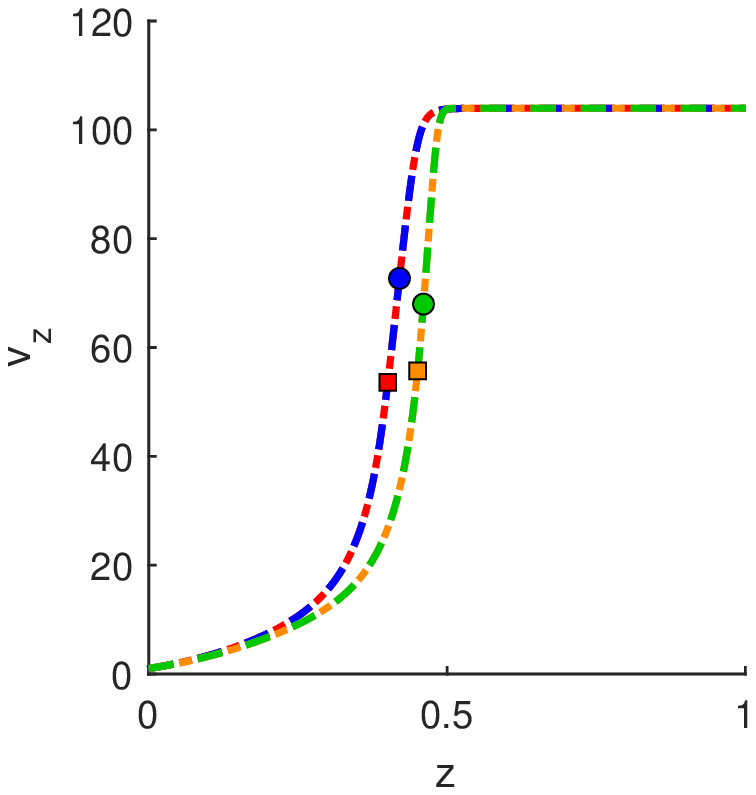}}
\hspace{0.5cm}
\subfigure{\includegraphics[width=0.3\textwidth]{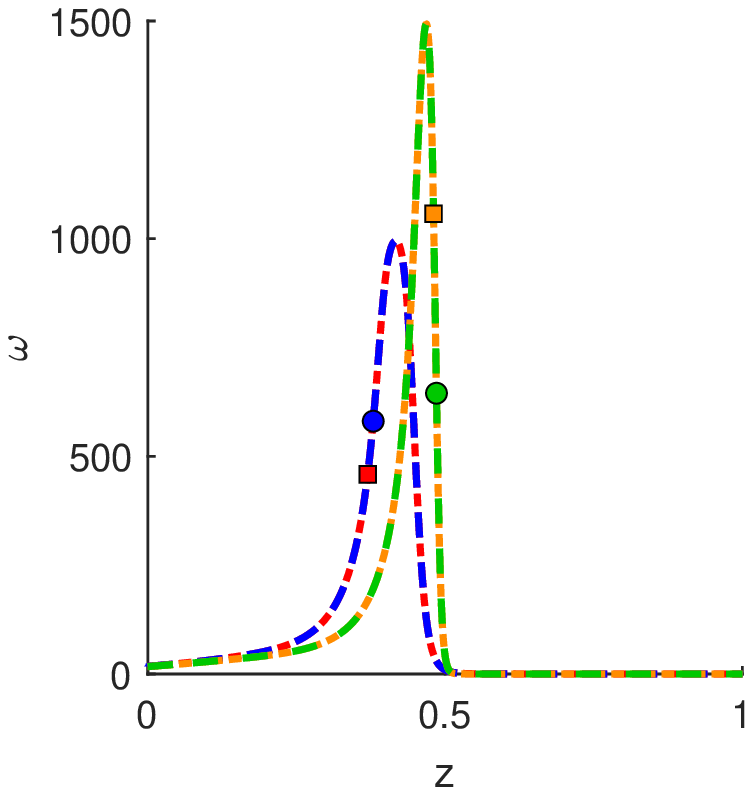}}
\hspace{0.5cm}
\subfigure{\includegraphics[width=0.3\textwidth]{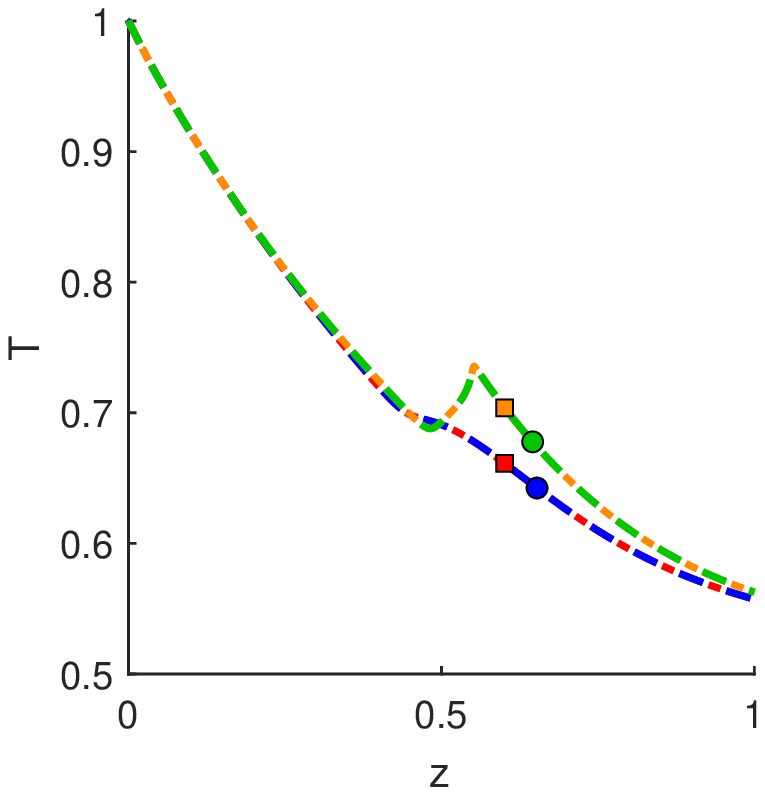}}\\
\subfigure{\includegraphics[width=0.3\textwidth]{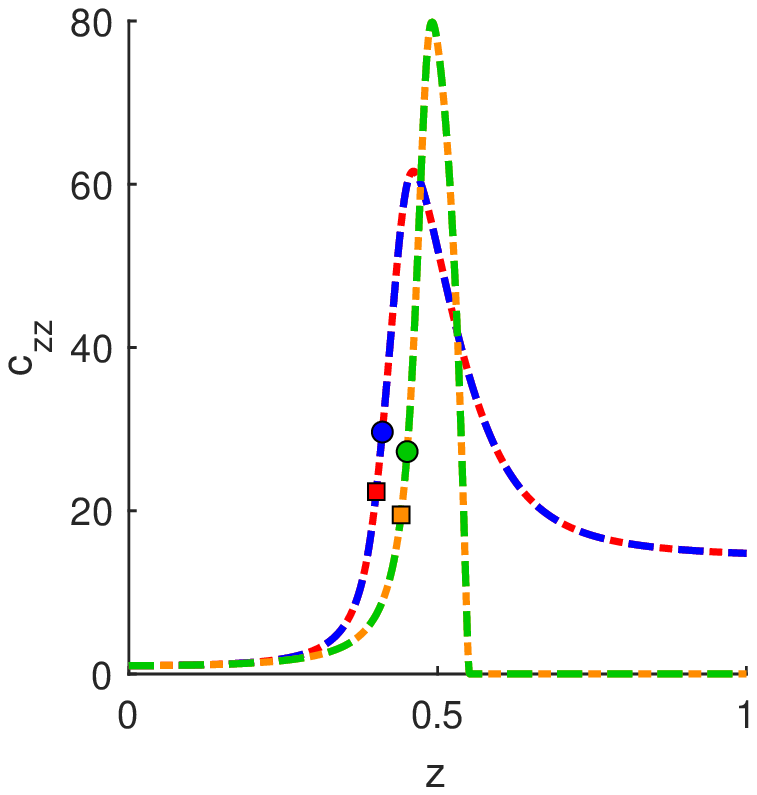}}
\hspace{0.5cm}
\subfigure{\includegraphics[width=0.3\textwidth]{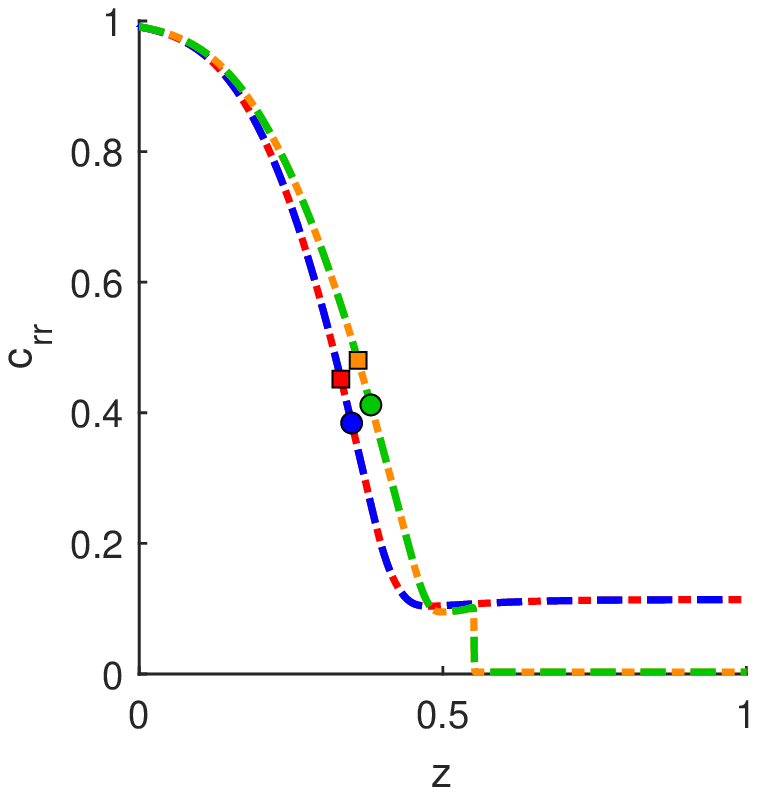}}
\hspace{0.5cm}
\subfigure{\includegraphics[width=0.3\textwidth]{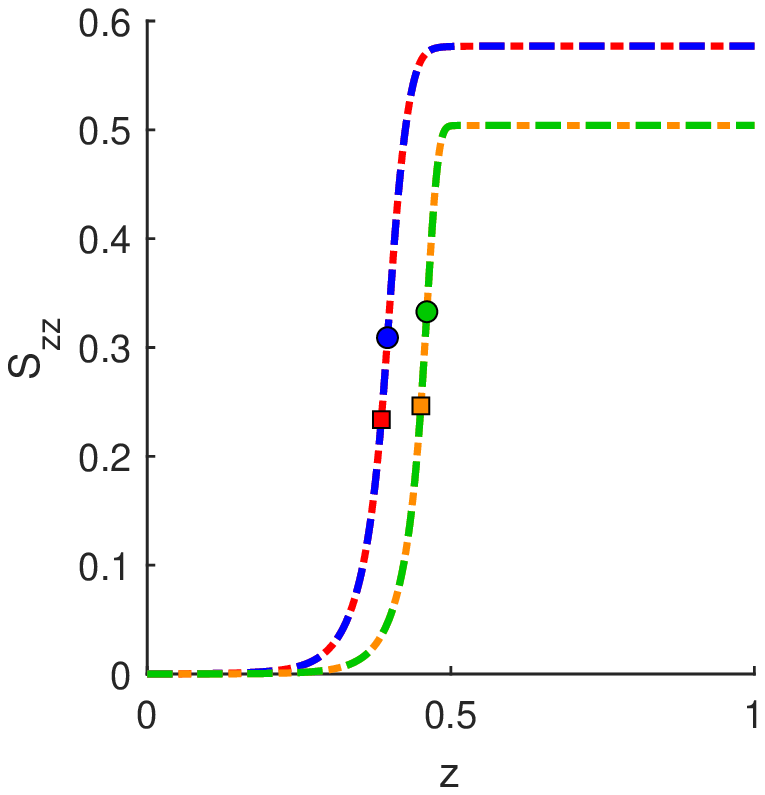}}\\
\subfigure{\includegraphics[width=0.3\textwidth]{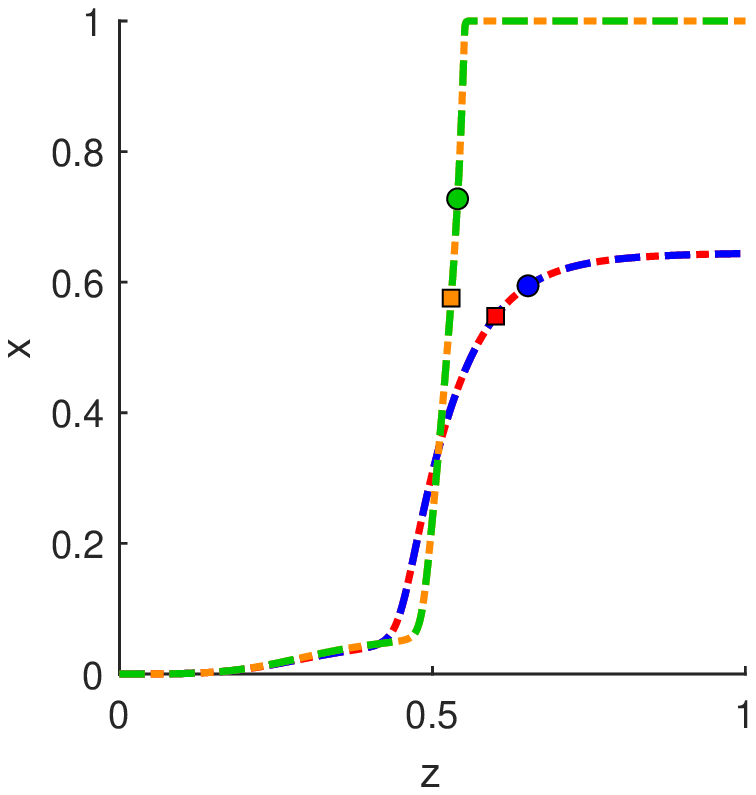}}
\hspace{0.5cm}
\subfigure{\includegraphics[width=0.3\textwidth]{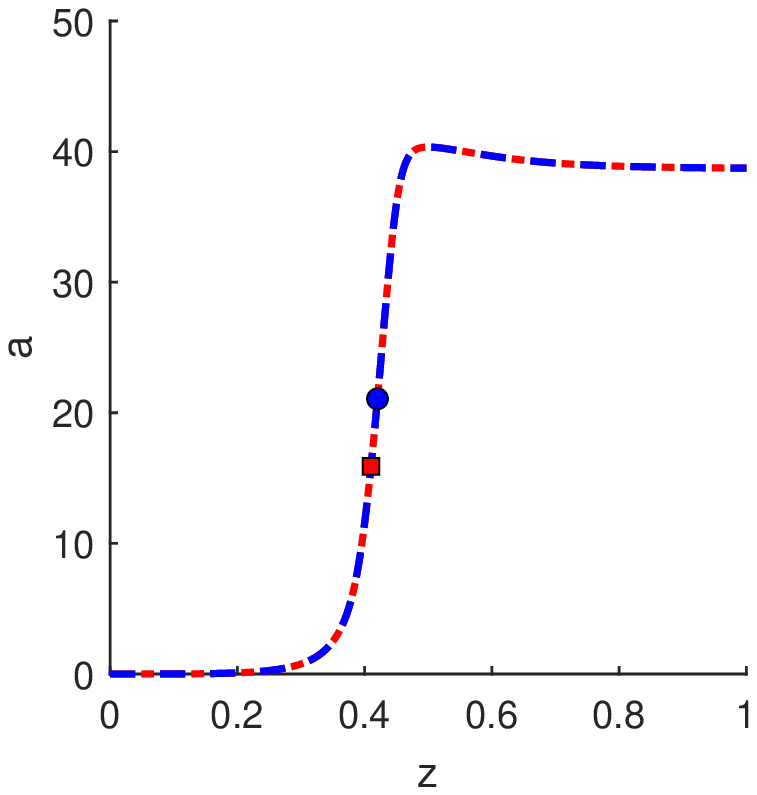}}
\subfigure{\includegraphics[width=0.33\textwidth]{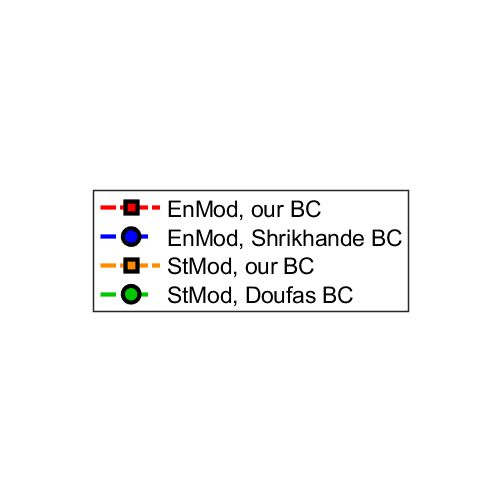}}
\caption{Fiber quantities obtained from stress-driven model (StMod) with our first-order and Doufas' boundary conditions as well as from energy-driven model (EnMod) with our and Shrikhande's boundary conditions, $P^\star=97.2095$.}
\label{fig_solution_results}
\end{figure}

In this section we investigate the impact of our asymptotically justified boundary conditions on the simulations and their performance. We discuss the effect on the solution behavior and highlight the computational improvements.

As test case we consider the high-speed melt spinning setup of Nylon-66 from Doufas et al.\ (labeled as S01 in   \cite{doufas_simulation_2000}), where rapid crystallization occurs along the spinline due to the cooling quench air. It has also served as a test example by Shrikhande et al.\ in \cite{shrikhande_modified_2006}.
All relevant physical, rheological and model parameters as well as the setup-specific reference values can be found in Appendix~\ref{appendix_test_case_S01}.
For the numerical treatment of the boundary value problems we employ a continuation-collocation scheme which has been successfully used in various fiber formation processes, see, e.g.,  \cite{arne_electrospinning_2017, arne_fluid-fiber-interactions_2011, wieland_efficient_2019, wieland_melt-blowing_2019}.
For further information on the numerics we refer to Appendix \ref{appendix_numerics}. We note that our collocation-continuation method has some advantages over the shooting approach in \cite{doufas_simulation_2000, shrikhande_modified_2006}. However, the computational improvements gained by our boundary conditions are independent of the used method.

All computations have been performed in MATLAB (version R2020a) on a system with an Intel Core i7-8665u CPU (4 cores, 8 threads) and 16GBytes of RAM.

\subsection{Results and boundary layers}

For the melt spinning setup the stress-driven model (System~\ref{system_final_model_doufas}) predicts a crystallinity value of one at the end of the spinline, whereas the (free stored) energy-driven model (System~\ref{system_final_model_shrikhande}) provides a significantly lower value at the take-up.
The more rapid crystallization process results in a steeper curvature of the velocity profile as well as a higher temperature increase due to crystallization. Our simulation results are in accordance with literature and presented below in the dimensionless fiber quantities. Our asymptotically justified boundary conditions lead to the same fiber behavior on the macro (fiber) scale as the original boundary conditions by Doufas et al.\ and Shrikhande et al., respectively, as visualized in Fig.~\ref{fig_solution_results} (using $P^\star=97.2095$ in \eqref{eq:bc_S}). Their effects are limited to a small region around the point of onset of crystallization. The differences are better recognizable in the derivatives of the fiber quantities, in particular of the acceleration $\omega$, the orientation stress $S_{zz}$ and the conformation tensor components $c_{zz}$, $c_{rr}$.

\begin{figure}[tbh]
\centering
\subfigure{\includegraphics[width=0.3\textwidth]{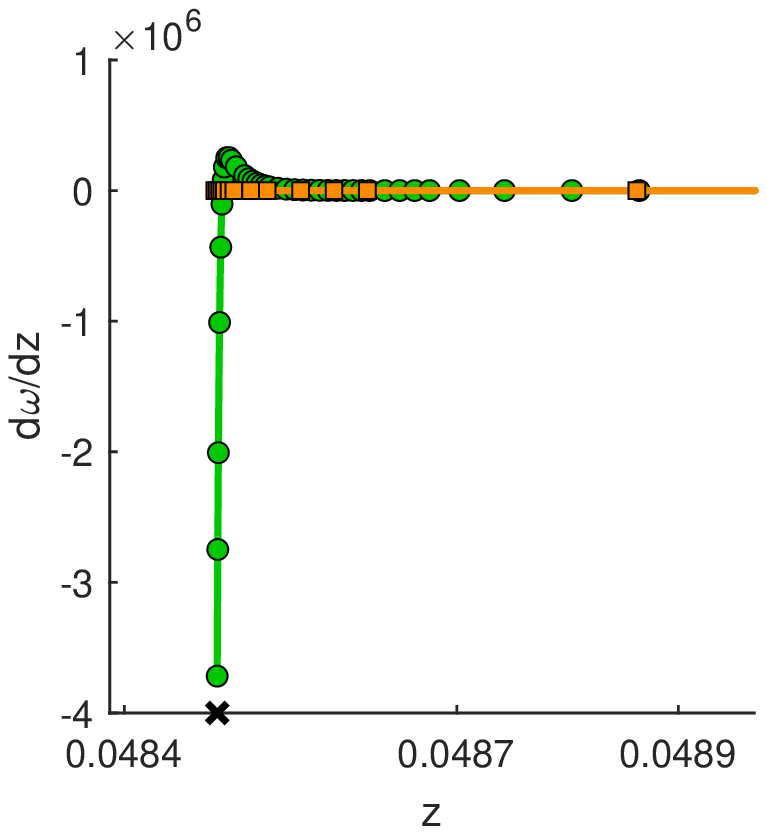}}
\hspace{1.5cm}
\subfigure{\includegraphics[width=0.3\textwidth]{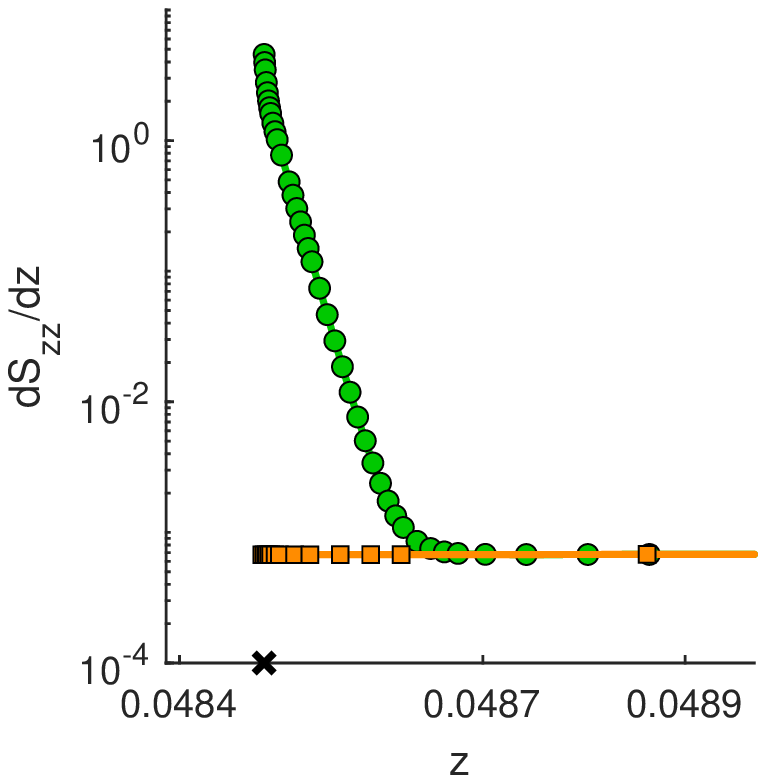}}\\
\subfigure{\includegraphics[width=0.3\textwidth]{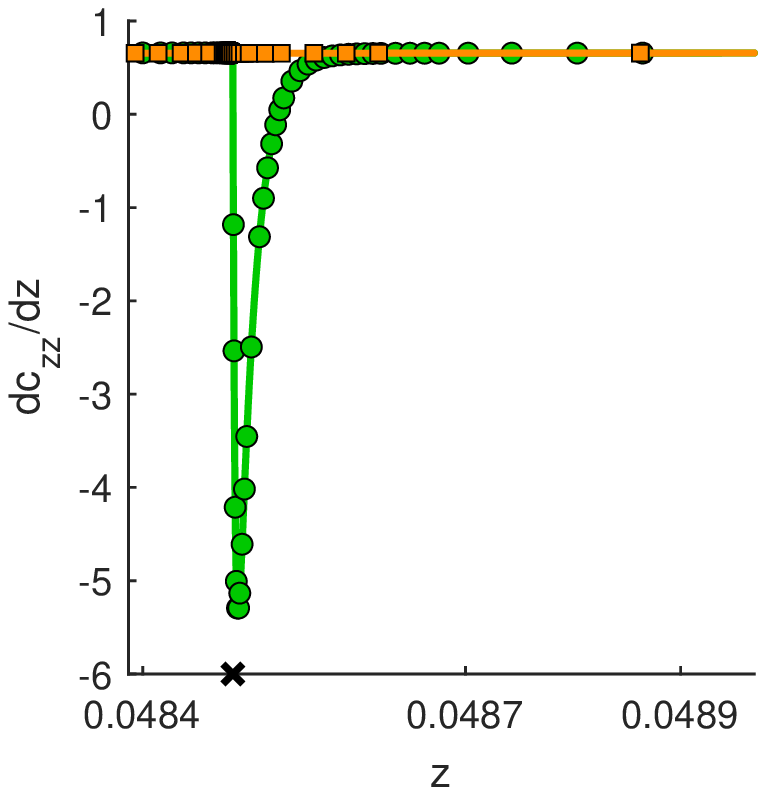}}
\hspace{1.6cm}
\subfigure{\includegraphics[width=0.3\textwidth]{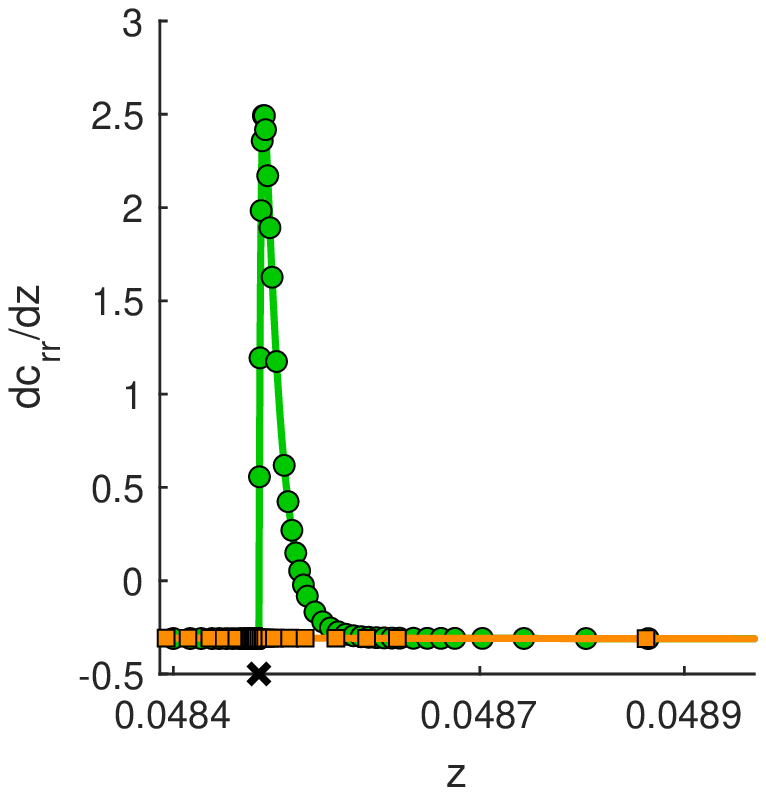}}\\
\caption{StMod with our first-order (orange $\square$) and Doufas' (green $\circ$) boundary conditions. Derivatives near the point of crystallization onset $z^\lozenge$ (black x).}
\label{fig_derivative_doufas_model}
\end{figure}

\begin{figure}[tbh]
\centering
\subfigure{\includegraphics[width=0.3\textwidth]{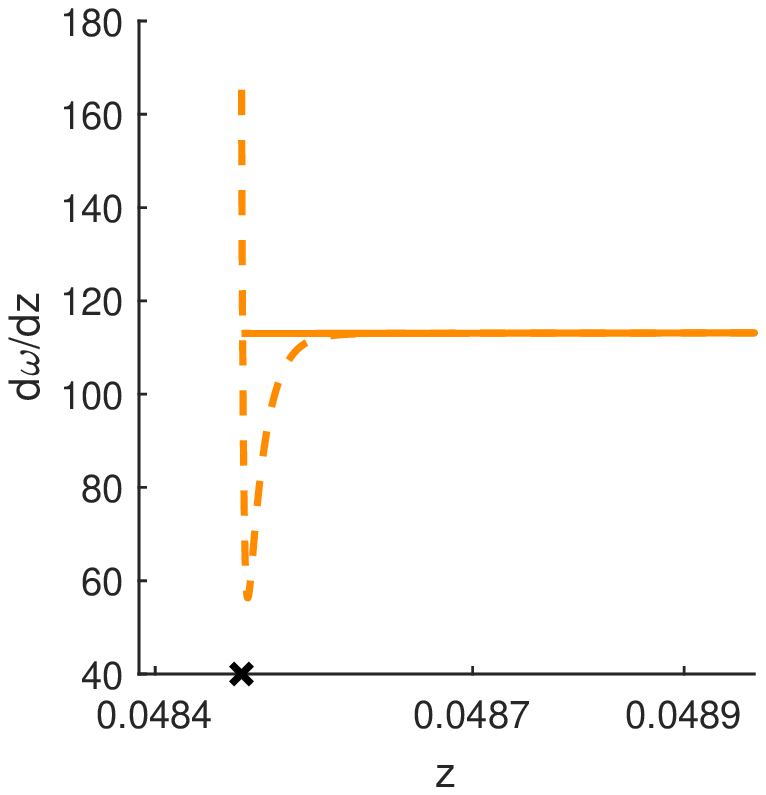}}
\hspace{1.5cm}
\subfigure{\includegraphics[width=0.3\textwidth]{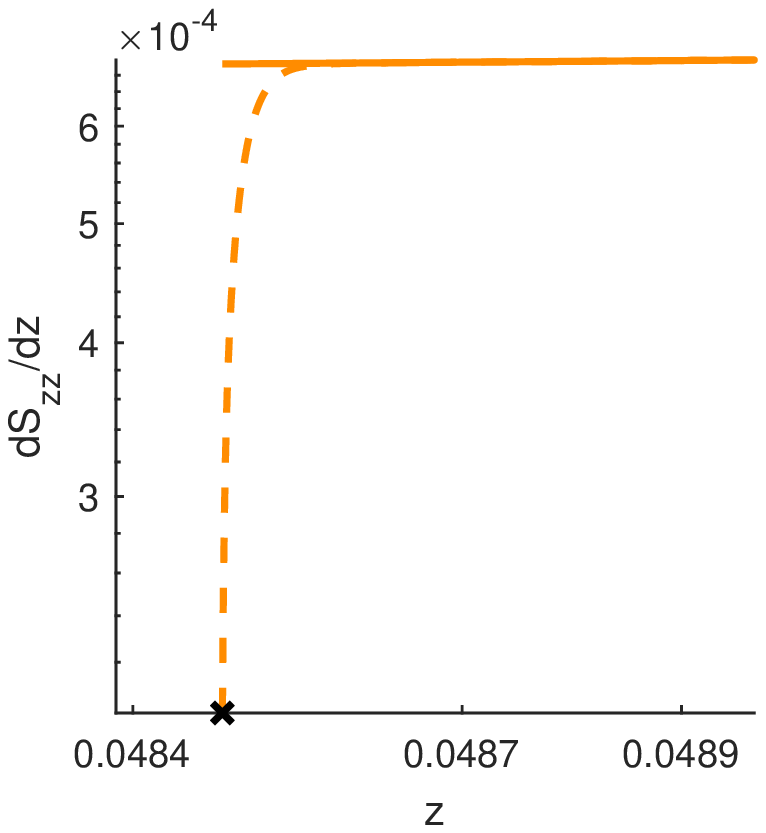}}\\
\subfigure{\includegraphics[width=0.3\textwidth]{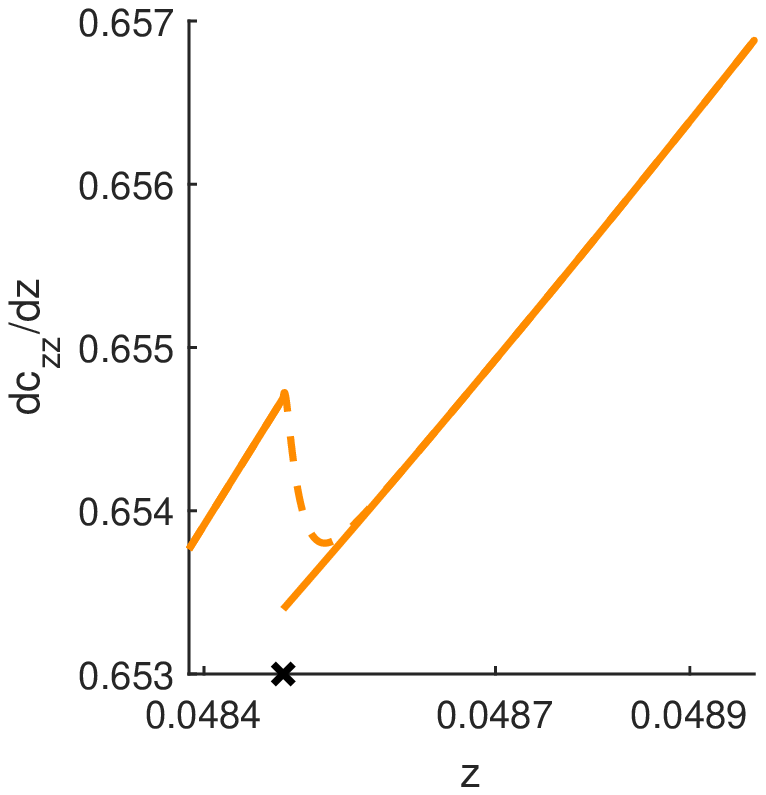}}
\hspace{1.6cm}
\subfigure{\includegraphics[width=0.3\textwidth]{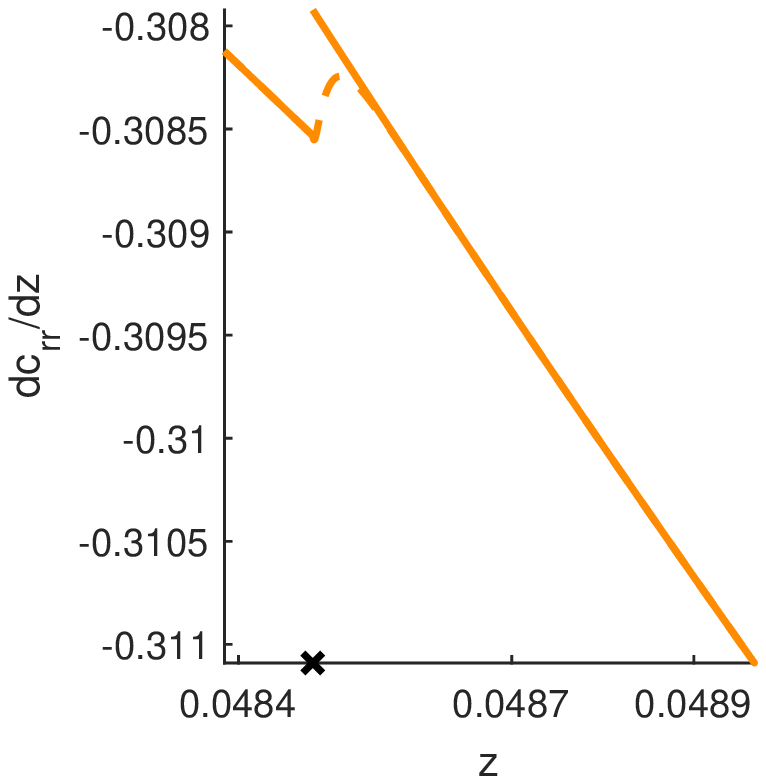}}\\
\caption{StMod with Case A (zeroth-order BC; dashed line) and Case B (first-order BC; solid line). Point of crystallization onset $z^\lozenge$ marked by cross.}
\label{fig_derivative_doufas_model_zerofirst}
\end{figure}

\subsubsection*{Stress-driven model}
For the stress-driven model (StMod) the (dimensionless) derivatives of $\omega$, $S_{zz}$, $c_{zz}$ and $c_{rr}$ are shown in Fig.~\ref{fig_derivative_doufas_model}.
In agreement with \cite{shrikhande_modified_2006}, the original boundary conditions by Doufas et al.\ lead to the formation of a strong boundary layer with large derivatives at the point of crystallization onset, here $z^\lozenge \approx 0.0485$. This is the interface location along the spinline where the fiber model switches between the describing non-crystallizing and crystallizing equations.
Both sets of our asymptotically justified boundary conditions (Case~A and Case~B) drastically reduce the layering and avoid the occurrence of high derivatives in the vicinity of the point of crystallization onset, as the values in this region are consistent to the values outside of it. The differences of Case~A  and Case~B are illustrated in Fig.~\ref{fig_derivative_doufas_model_zerofirst}.
Case~A is related to a zeroth-order asymptotics in $\delta$. It uses a continuous initialization of $\omega$ and $S_{zz}$ at the interface with respect to the states before onset of crystallization. Consequently, the quantities that exists along the entire spinline (e.g., the conformation tensor components) are smooth, i.e., they have continuous derivatives. The quantities $\omega$ and $S_{zz}$ themselves imprint a slight layering in the derivatives due to the switching from the non-crystallizing equations $\delta=\tn{Da}_\tn{I} = 0$ to the crystallizing equations $\delta \neq 0$, $\tn{Da}_\tn{I} \neq 0$. Case~B, in contrast, accounts for the model switching by handling the point of crystallization onset as a non-smooth interface. Considering $\omega$ and $S_{zz}$ exclusively as quantities of the crystallizing equations, the first-order asymptotics provides initializations for which $\omega$ and $S_{zz}$ have constant derivatives without any layer formation, i.e., the proposed boundary conditions match very well the global behavior. But in this case the other quantities compensate the model switching with a jump in the derivatives at the interface (cf.\ $\text{d}c_{zz}/\text{d}z$, $\text{d}c_{rr}/\text{d}z$ in Fig.~\ref{fig_derivative_doufas_model_zerofirst}). From the mathematical point of view, the two cases demonstrate the possible treatment of the interface, i.e., smooth versus non-smooth. They are equal in importance.

In the considered melt spinning setup the crystallinity $x$ of the stress-driven model approaches one and its derivative zero. As discussed in Remark~\ref{rem:2}, the numerics may break down due to raising singularities. Doufas et al.\ \cite{doufas_simulation_2000-1} overcame this problem by introducing a stopping criterion and performing a model switching. The price to pay is a non-differentiable solution at the switching point and an error-prone numerics. We propose a regularization to ensure smooth solutions and good solver properties. In particular, our reformulation of System~\ref{system_final_model_doufas} as a boundary value problem with fixed interface and regularization allows its very robust and efficient simulation as one cohesive model without any  "switch-like conditions", see Appendix~\ref{subsec:System2} for details to the numerical treatment and Section~\ref{sec:solver} for the solver performance.

\begin{figure}
\centering
\hspace{0.4cm}
\subfigure{\includegraphics[width=0.3\textwidth]{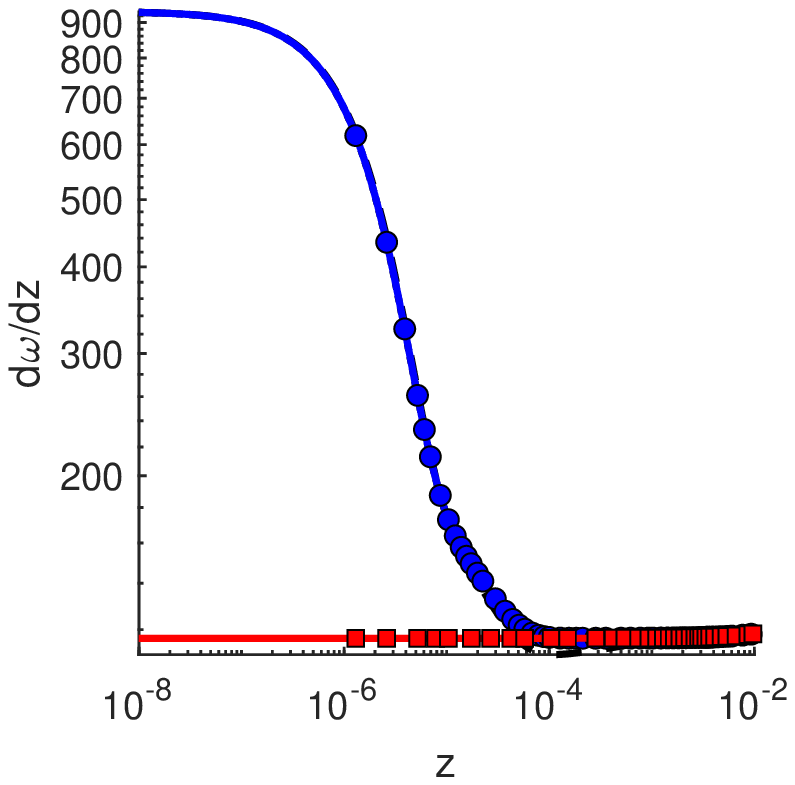}}
\hspace{0.6cm}
\subfigure{\includegraphics[width=0.3\textwidth]{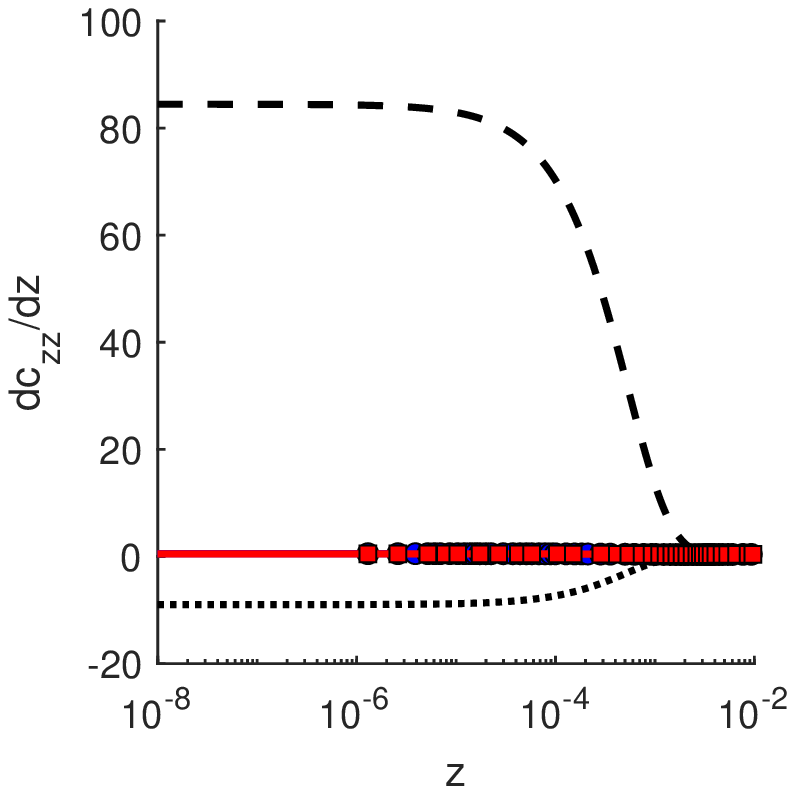}}
\subfigure{\includegraphics[width=0.29\textwidth]{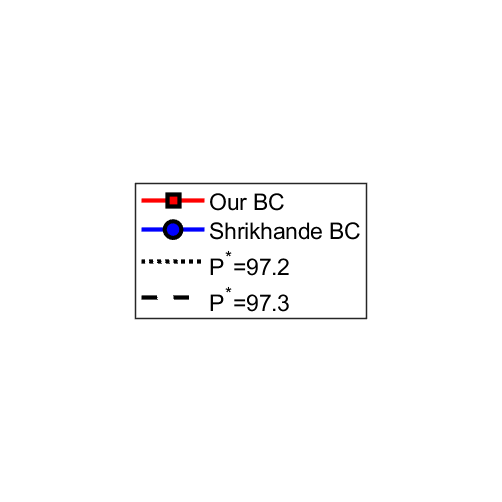}}\\
\subfigure{\includegraphics[width=0.3\textwidth]{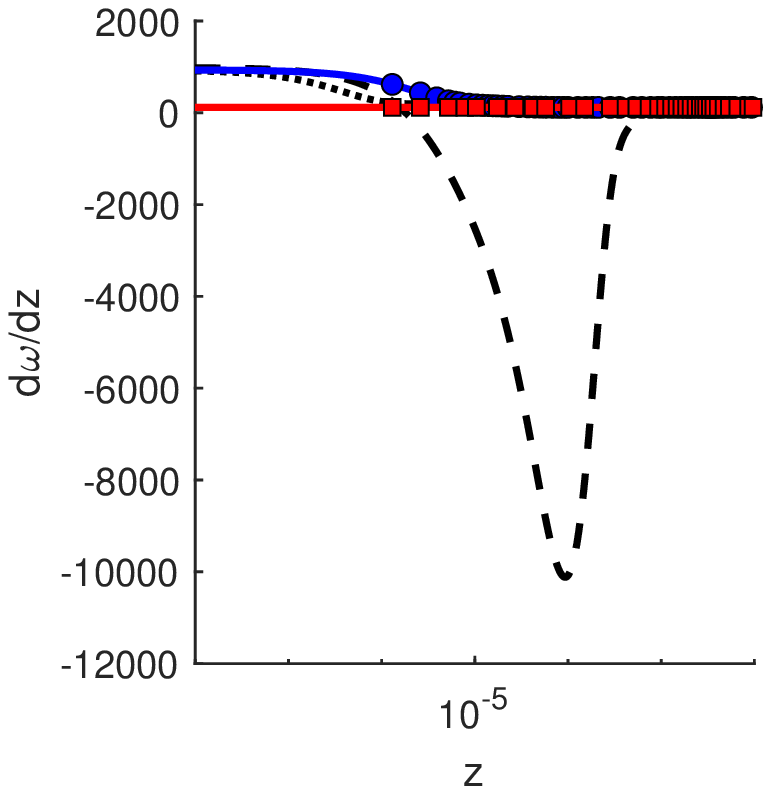}}
\hspace{1cm}
\subfigure{\includegraphics[width=0.3\textwidth]{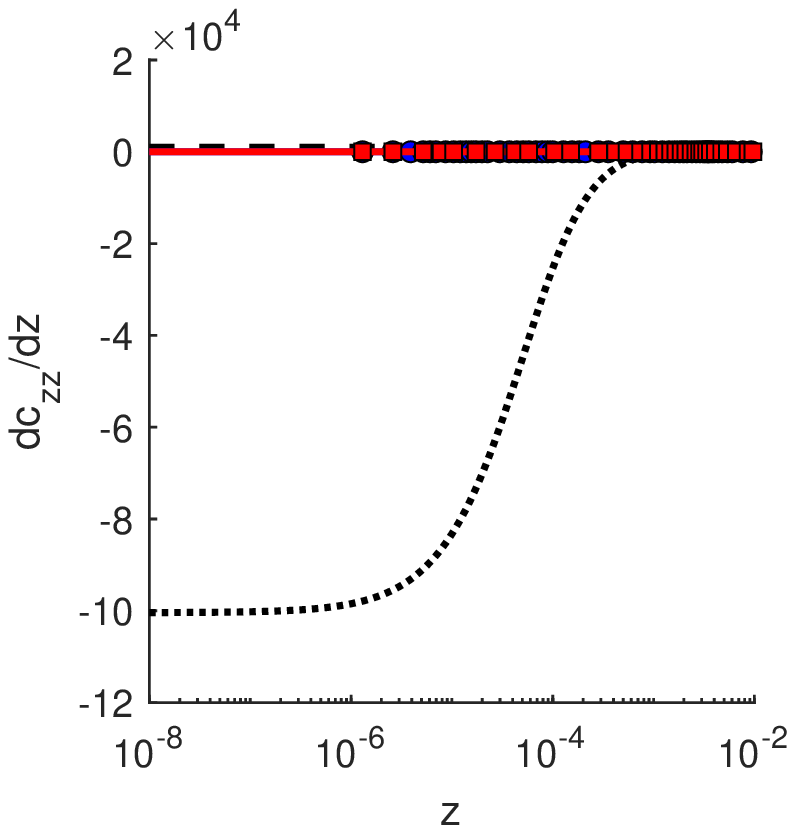}}
\subfigure{\includegraphics[width=0.29\textwidth]{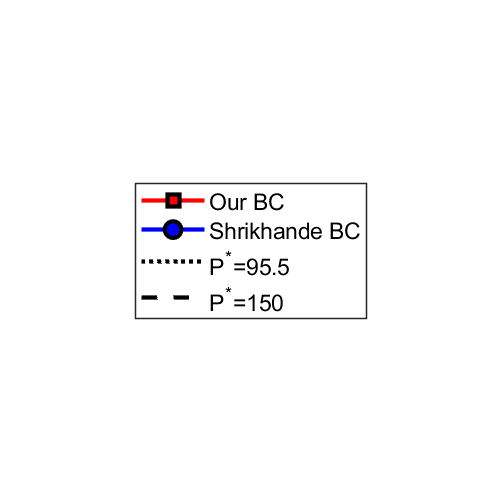}}\\
\caption{EnMod with our first-order (red $\square$) and Shrikhande's (blue $\circ$) boundary conditions with $P^*=97.2095$ as well as with different variations of the parameter $P^*$.
The grid values of the numerical solver are indicated by the markers, the lines are obtained from a piecewise cubic Hermite interpolation of the solution quantities and a subsequent evaluation of the right hand side functions.
Note that there is another grid point at $z=0$, which cannot be plotted on the logarithmic scale.}
\label{fig_derivative_shrik_model}
\end{figure}

\begin{figure}
\centering
\begin{tabular}{c@{\hspace{1cm}}cc}
\subfigure{\includegraphics[width=0.3\textwidth]{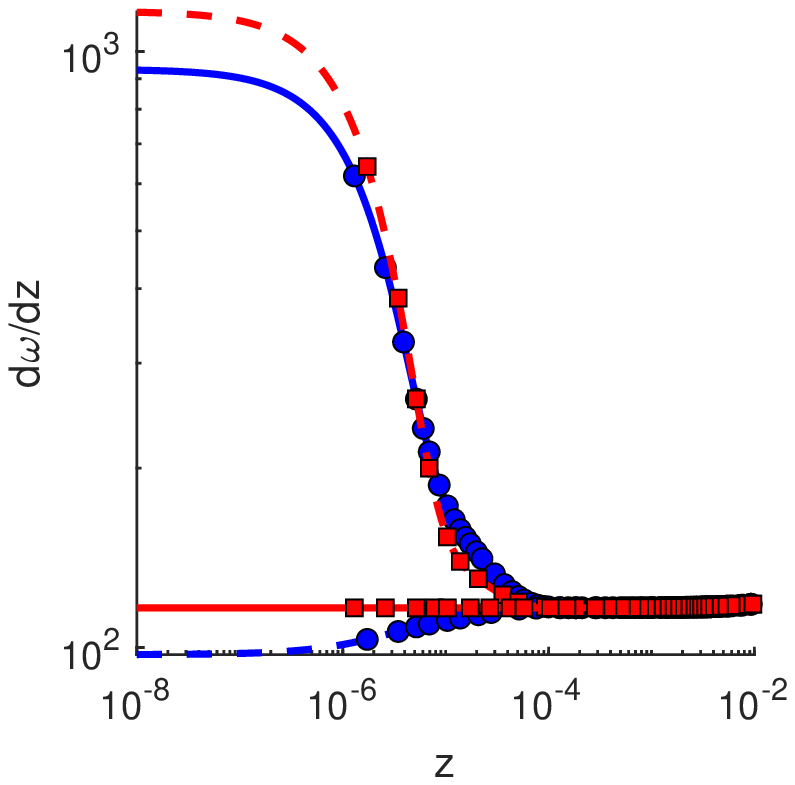}}&
\subfigure{\includegraphics[width=0.3\textwidth]{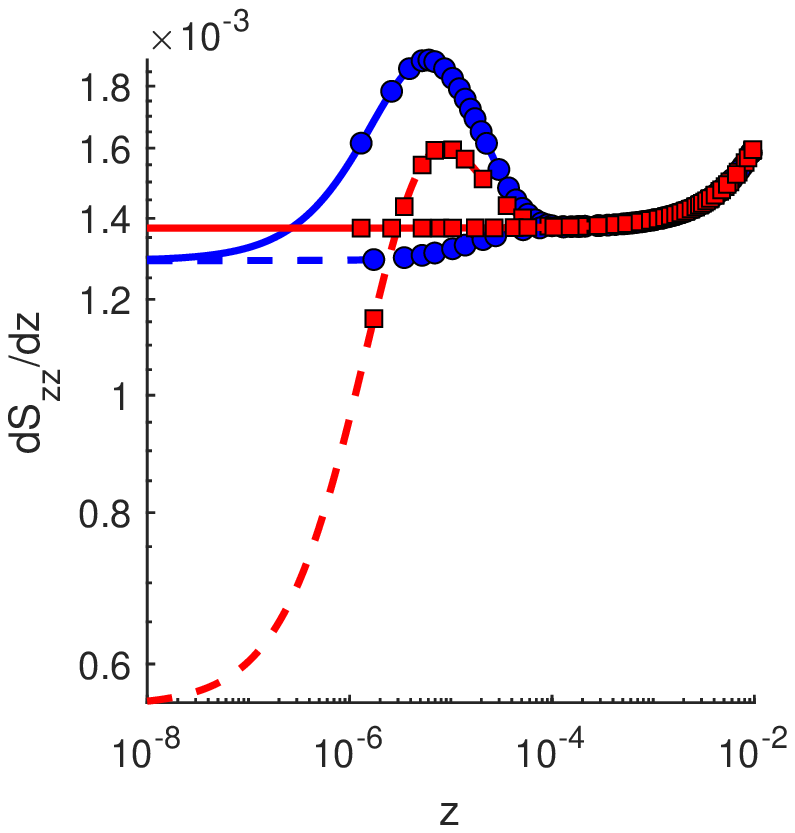}}&
\multirow{2}{*}[1.8cm]{
    \subfigure{\includegraphics[width=0.29\textwidth]{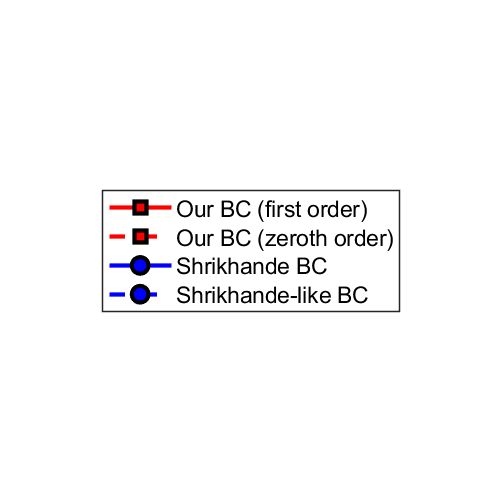}}
}\\
\subfigure{\includegraphics[width=0.3\textwidth]{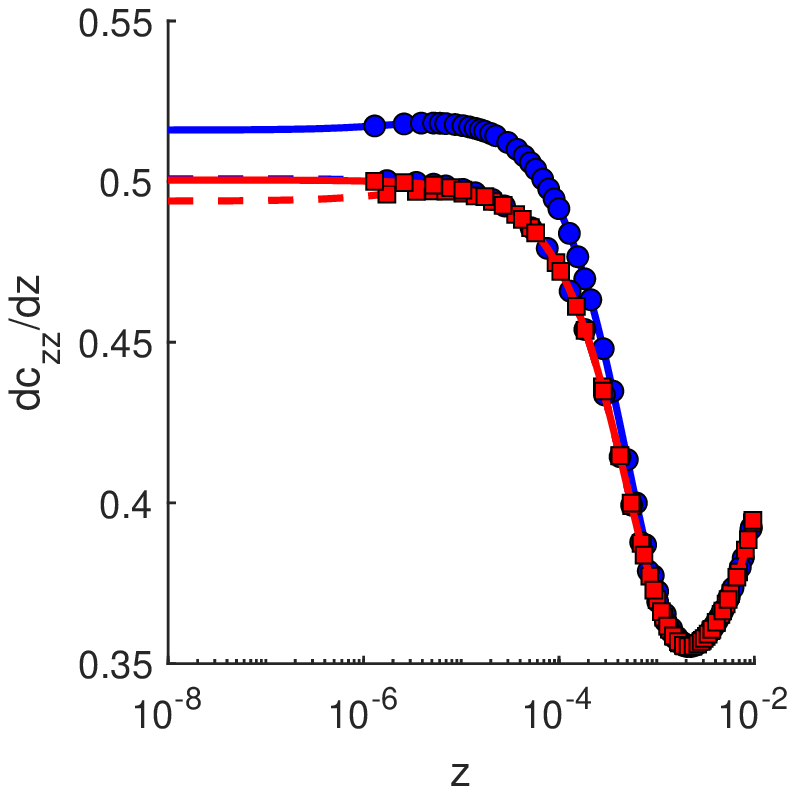}}&
\subfigure{\includegraphics[width=0.3\textwidth]{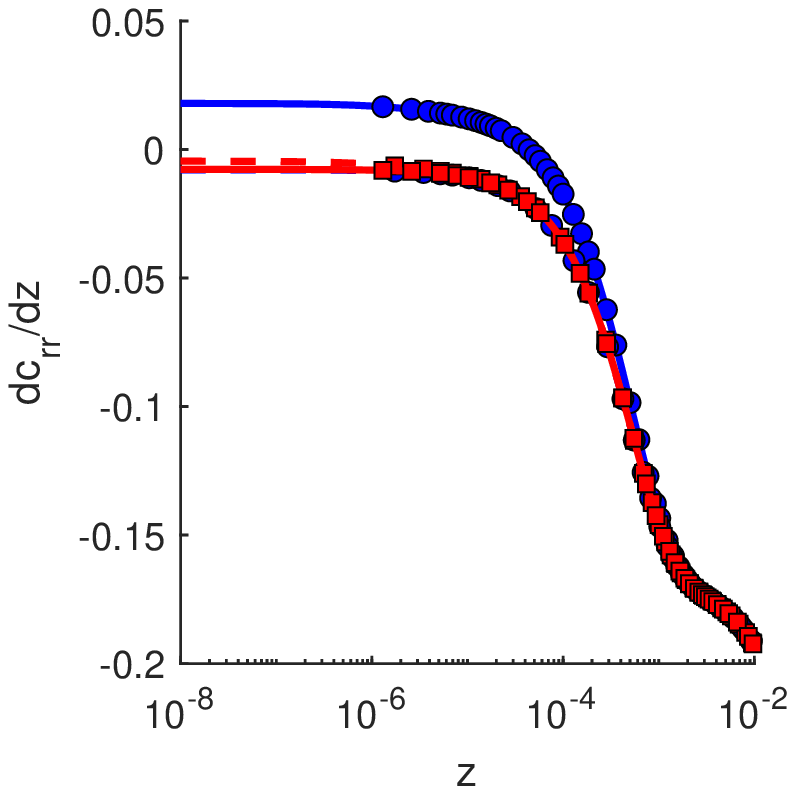}}&
\end{tabular}
\caption{EnMod with our first-order and zeroth-order (red $\square$; solid and dashed line) as well as Shrikhande's and Shrikhande-like (blue $\circ$; solid and dashed line) boundary conditions.}
\label{fig_derivative_extended}
\end{figure}

\newpage
\subsubsection*{Energy-driven model} 
In the energy-driven model (EnMod) the point of crystallization onset is located at the inlet $z=0$. Figure~\ref{fig_derivative_shrik_model} illustrates the influence of the parameter $P^\star$ that has to be chosen in the original boundary conditions by Shrikhande et al.\ \eqref{eq:bc_S}. Already small moderate variations in $P^\star$ involve pronounced changes in the boundary layer behavior. If the parameter is chosen poorly, this leads to an uncontrollable layering at the inlet due to the lack of a suitable boundary condition with respect to the conformation tensor components. In an attempt to minimize the occurring boundary layer by parameter tuning we found that $P^\star=97.2095$ yields the best results for Shrikhande's boundary conditions in this special test case (cf.\ original work \cite{shrikhande_modified_2006}). But even these boundary conditions produce a significant layering in the considered quantities. The behavior can be improved by the Shrikhande-like conditions \eqref{eq:bc_alternative} (cf.\ Remark~\ref{rem:shriklike_bc}), where the free parameter  $P^\star$ is determined by the viscous relation for the conformation tensor components, i.e., $c_{zz}(0)+2c_{rr}(0)=3$, see Fig.~\ref{fig_derivative_extended}. As expected, the Shrikhande-like conditions also lead to better results than our asymptotically justified zeroth-order conditions, as they contain some additional first-order terms in $\delta$. However, the complete elimination of artificial layering (especially in $\omega$ and $S_{zz}$) is guaranteed only by our first-order boundary conditions. They prescribe the boundary conditions consistently up to an error of $\mathcal{O}(\delta^2)$, here $\delta=0.02$.

\newpage
\subsection{Solver performance}\label{sec:solver}
The elimination of artificial layering by means of our asymptotically justified boundary conditions has an impact on the computational performance of the numerical solvers. For prescribed error tolerances the required resolution significantly reduces. The number of grid points required in the boundary region as well as for the entire fiber are listed in Tab.\ \ref{tab_grid_comparison}.
For the stress-driven model the number of grid points is reduced by about 55\% in the boundary region compared to the conditions by Doufas et al., for the energy-driven model by about 35\% compared to the conditions by Shrikhande et al. The total number of grid points is less in both model setups.
  
As another indication for an improvement to the numerics we point to the run time of the solver. The total run time for the stress-driven model is reduced by approximately 
15\%, from 367 seconds using Doufas' boundary conditions to 311 seconds using our boundary conditions. For the energy-driven model there is no notable speed-up, when using the best fit parameter $P^\star=97.2095$.
However, note that this parameter must be readjusted for each test case which leads to a massive overhead in computing time. Moreover, the parameter tuning makes the numerics error-prone.

\begin{table}[tb]
\centering
\begin{tabular}{|l r r|}
\hline
									& boundary region & entire spinline\\
Interval length									& 1\%	& 100\%	\\
\hline
\textbf{Stress-driven model} && \\
Case A (zeroth-order boundary conditions)	& 26	& 517	\\
Case B (first-order boundary conditions)		& 26	& 517	\\
Doufas' boundary conditions			& 58	& 558	\\
\hline
\textbf{Energy-driven model} && \\
First-order boundary conditions		& 35	& 318	\\
Zeroth-order boundary conditions	& 36	& 305	\\
Shrikhande's boundary conditions		& 57	& 334 	\\
Shrikhande-like boundary conditions	& 37	& 316 	\\
\hline
\end{tabular}
\vspace*{0.2cm}
\caption{Required resolution (number of grid points) using our numerical solver with relative error tolerance $\mathcal{O}(10^{-5})$ and absolute error tolerance  $\mathcal{O}(10^{-10})$.}
\label{tab_grid_comparison}
\end{table}

\section{Conclusion}
The focus of this paper was on the establishment of appropriate boundary conditions for the viscoelastic two-phase fiber model class by Doufas et al.\ \cite{doufas_simulation_2000-1} and Shrikhande et al.\  \cite{shrikhande_modified_2006}. Our asymptotic analysis revealed that the model system of ordinary differential equations reduces to a system of differential-algebraic equations in the limit of a vanishing semi-crystalline relaxation time and that boundary layers might arise due to a singular perturbation. By deducing asymptotically justified boundary conditions for the point of crystallization onset, we formulated regularly perturbed boundary value problems. In the application for high-speed melt spinning of Nylon-66 our simulation results are in good agreement with literature. The effects of our boundary conditions on the overall solution behavior are restricted to a small region near the crystallization onset, while the achieved computational improvements are remarkably large.  Since the occurrence of large boundary layers and artificial discontinuities are prevented, the resolution (number of grid points) required for the same accuracy is reduced. This is accompanied by less computational effort and faster run times. Moreover, the numerics becomes robust, as ambiguities and parameter tunings are avoided. The achieved performance enhancement enables design and optimization of industrial spinning processes that feature thousands of fibers and two-way coupled fiber-air-interactions.

\appendix
\renewcommand{\theequation}{\Alph{section}.\arabic{equation}}
\renewcommand{\thetable}{\Alph{section}.\arabic{table}}
\renewcommand{\thefigure}{\Alph{section}.\arabic{figure}}
\setcounter{equation}{0} \setcounter{figure}{0} \setcounter{table}{0}

\section{Asymptotic Derivation of Boundary Conditions} \label{appendix_asymptotic}
The task to find appropriate closing boundary conditions for the viscoelastic fiber model class can be embedded in the following general asymptotic consideration.
Given a $\delta$-perturbed system of first-order ordinary differential equations on $I=(0,1)$ of the form
\begin{subequations}
\begin{align}
\deriv{\bs{x}} &= \bs{F}(\bs{x},\bs{y};\delta), \qquad \qquad \qquad 0 < \delta \ll 1, \label{eqApp:ode_dynamic_general} \\
\delta \,\deriv{\bs{y}} &= \bs{G}(\bs{x},\bs{y};\delta), \label{eqApp:ode_algebraic_general} \\
\bs{0} &= \bs{H}(\bs{x}(0),\bs{x}(1))\in \mathbb{R}^N \label{eqApp:boundaryvaluefunction}
\end{align}
\end{subequations}
for the unknowns $\bs{x}:I\rightarrow \mathbb{R}^N$ and $\bs{y}:I\rightarrow \mathbb{R}^M$. The question is how to choose the remaining $M$ boundary conditions to obtain a regularly perturbed boundary value problem.

In the asymptotic limit $\delta = 0$ the equations degenerate to a differential-algebraic system, where \eqref{eqApp:ode_algebraic_general} become algebraic relations. In case of a singular perturbation, the remaining boundary conditions to be posed are not consistent to the algebraic relations, which causes the raising of boundary layers or even the occurrence of discontinuities in the solution for $\delta\geq 0$. A regular perturbation preventing this can be derived as follows.

Assuming a regular expansion of the unknowns in $\delta$, i.e., $\bs{x}=\sum_{i} \delta^{i} \bs{x}^{(i)}$ and $\bs{y}=\sum_{i} \delta^{i} \bs{y}^{(i)}$, we formally expand \eqref{eqApp:ode_algebraic_general} in $\delta$ and search for boundary conditions that satisfy \eqref{eqApp:ode_algebraic_general} up to an error of $\mathcal{O}(\delta^{p+1})$ for $p\in \mathbb{N}_0$. We refer to them as $p$th-order boundary conditions. They yield regular perturbations of $p$th order. In zeroth order, we find from \eqref{eqApp:ode_algebraic_general}
\begin{align}\label{eq:A1}
\bs{0}= \bs{G}(\bs{x}^{(0)},\bs{y}^{(0)};0),
\end{align}
consequently, the algebraic relations $0= \bs{G}(\bs{x},\bs{y};0)$ themselves give zeroth-order boundary conditions. For first-order boundary conditions we proceed from \eqref{eqApp:ode_algebraic_general} in the form
\begin{align}\label{eq:A2}
\delta \deriv{\bs{y}^{(0)}} +\mathcal{O}(\delta^2)= \bs{G}(\bs{x},\bs{y};\delta)
\end{align}
and derive a $\delta$-consistent expression for $\text{d}{\bs{y}^{(0)}}/\text{d}z$.
Differentiating \eqref{eq:A1} with respect to $z$ and using \eqref{eqApp:ode_dynamic_general} yields
\begin{align*}
\bs{0}=\partial_{\bs{x}} \bs{G}(\bs{x}^{(0)},\bs{y}^{(0)};0) \cdot \bs{F}(\bs{x}^{(0)},\bs{y}^{(0)};0) + \partial_{\bs{y}} \bs{G}(\bs{x}^{(0)},\bs{y}^{(0)};0) \cdot \deriv{\bs{y}^{(0)}}.
\end{align*}
Assuming the invertibility of $\partial_{\bs{y}}\bs{G}$, this involves an explicit expression for  $\text{d}{\bs{y}^{(0)}}/\text{d}z$ that keeps its accuracy in leading order even when in the function arguments the zeroth-order quantities are replaced by the full expansions and the $\delta$-dependence of $\bs{F}$ is considered. Inserting it in \eqref{eq:A2}, the relations
\begin{align*} \label{eqApp:asympBC_general}
\bs{0} = \delta \, \partial_{\bs{x}} \bs{G}\left(\bs{x},\bs{y};0\right) \cdot \bs{F}\left(\bs{x},\bs{y};\delta\right) + \partial_{\bs{y}} \bs{G}\left(\bs{x},\bs{y};0\right) \cdot \bs{G}\left(\bs{x},\bs{y};\delta\right)
\end{align*}
give first-order boundary conditions.

\section{Closure approximation for semi-crystalline phase} \label{appendix_embedding}
The closure approximations $\closureFz(S_{zz})$ and $\closureFr(S_{zz})$ for the semi-crystalline phase in \eqref{eq:model}  result from microstructural considerations (see \cite{doufas_simulation_2000-1}). They can be expressed as polynomials in the orientational tensor component $S_{zz}$, i.e.,
\begin{align*}
\closureFz(S_{zz})		&= -\frac{81}{8}S_{zz}^5 + \frac{675}{56}S_{zz}^4 - \frac{36}{35}S_{zz}^3 - \frac{9}{10}S_{zz}^2 + \frac{11}{14}S_{zz} + \frac{2}{15}, \\
\closureFr(S_{zz})		&= \phantom{-}\frac{81}{16}S_{zz}^5 - \frac{675}{112}S_{zz}^4 + \frac{18}{35}S_{zz}^3 + \frac{9}{20}S_{zz}^2 + \frac{5}{14}S_{zz} - \frac{1}{15}.
\end{align*}

\section{Melt spinning of Nylon-66: Closing models and parameters}\label{appendix_test_case_S01}

The test case, the melt spinning setup of Nylon-66, comes from \cite{doufas_simulation_2000}, see also \cite{shrikhande_modified_2006}. It is characterized by a high take-up velocity (large draw ratio) and a rapid crystallization due to the cooling quench air. The closing models and parameters used for the two-phase fiber models are briefly summarized in the following.

\subsubsection*{Material properties}
The models for the fiber density $\rho$, dynamic viscosity $\mu$, specific heat capacity $C_\mathrm{p}$ and specific latent heat of crystallization $\Delta H_\mathrm{f}$ of a Nylon-66 fiber are taken from \cite{doufas_simulation_2000-1, shrikhande_modified_2006}. The fiber density $\rho$ is considered to be constant;
\begin{alignat*}{2}
\mu(T)	&= \mu_{\mathrm{ref}} \exp\left(\frac{E_\mathrm{A}}{E_1} \frac{T_1 + T_2 - T}{T}\right), \\
C_\mathrm{p}(T,x)		&= C_\mathrm{p}^{\mathrm{cr}}(T) \,x \Phi_{\infty} + C_\mathrm{p}^{\mathrm{am}}(T) \,\left(1-x \Phi_{\infty}\right), \\
& \,\, \quad C_\mathrm{p}^{(\mathrm{cr})}(T)	\,\,= C_{\mathrm{s}1} + C_{\mathrm{s}2} (T-T_1), \\
& \,\, \quad C_\mathrm{p}^{(\mathrm{am})}(T)	= C_{\mathrm{l}1} + C_{\mathrm{l}2} (T-T_1), \\
\Delta H_\mathrm{f}(T)	&= \Delta H_\mathrm{ref} + (C_{\mathrm{l}1}-C_{\mathrm{s}1})(T-T_1) + (C_{\mathrm{l}2}-C_{\mathrm{s}2}) \frac{(T-T_1)^2}{2}.
\end{alignat*}
For the values of referential viscosity $\mu_{\mathrm{ref}}$ [Pa s], activation energy $E_\mathrm{A}$ [J/mol], referential heat of crystallization $\Delta H_{\mathrm{ref}}$ [J/kg] and ultimate degree of crystallization $\Phi_{\infty}$ see Table~\ref{tab_nylon_physical_parameters}, the other parameters are 
\begin{alignat*}{3}
&  T_1 = 273.15 \;\tn{K}, 		&& T_2 = 280 \; \tn{K}, && E_1 = 4599.05 \; \tn{J/mol},\\
& C_{\mathrm{s}1} = 1.255 \cdot 10^{3}\; \tn{J/(kg K)}, 		&& C_{\mathrm{s}2} = 8.368 \; \tn{J/(kg K}^2), \\
& C_{\mathrm{l}1} = 2.092 \cdot 10^{3} \; \tn{J/(kg K}), \qquad && C_{\mathrm{l}2} = 1.946 \;  \tn{J/(kg K}^2).
\end{alignat*}
The relaxation time $\lambda$ is described with a constant shear modulus $G$ as
 \begin{align*}
 \lambda(T) = \frac{\mu(T)}{G}, \qquad \quad
\deriv{\lambda} = - \lambda \frac{E_\mathrm{A}}{E_1} \frac{T_1+T_2}{T^2} \deriv{T}.
\end{align*}
The chosen values for the remaining model parameters that are associated to the two phases and the crystallization are listed in Table~\ref{tab_shrikhande_model_parameters}.

\subsubsection*{Aerodynamic drag and heat transfer}
The density $\rho_\mathrm{a}$, dynamic viscosity $\mu_\mathrm{a}$ and thermal conductivity $k_\mathrm{a}$ of the quench air are modeled as dependent on the temperature of fiber $T$ and air $T_\mathrm{a}$, according to \cite{doufas_simulation_2000-1, jeon_modeling_2008},
\begin{alignat*}{2}
&\rho_\mathrm{a}(T) 	&&= \frac{2P}{(T + T_\mathrm{a})R_s},\\
&\mu_\mathrm{a}(T)	&&= \beta_1 \frac{(0.5(T+T_\mathrm{a}))^{1.5}}{0.5(T+T_\mathrm{a}) + \beta_2}, \\
&k_\mathrm{a}(T)		&&= \beta_3 \, (0.5(T+T_\mathrm{a}) )^{0.866}
\end{alignat*}
with pressure $P = 1 \; \tn{atm} = 101325 \; \tn{Pa}$, specific gas constant of dry air $R_s = 287.05$ J/(kg K) and the parameters
\begin{equation*}
\beta_1 = 1.446 \cdot 10^{-6} \; \tn{Pa\;s K}^{-0.5},	\qquad \beta_2 = 113.9 \; \tn{K}, \qquad \beta_3 = 1.880 \cdot 10^{-4} \; \tn{W/(m$\tn{K}^{1.866}$)}.
\end{equation*}

\begin{table}[tb]
\centering
\begin{tabular}{|l l l l|}
\hline
\multicolumn{4}{|l|}{\textbf{Physical and rheological parameters}}						\\
Description								& Symbol			& Value		& Unit			\\
\hline
Density									& $\rho$			& 1106 		& kg/$\tn{m}^3$ \\
Referential viscosity at 280$^\circ$C	& $\mu_\text{ref}$		& 126.05	& Pa s			\\
Activation energy						& $E_\text{A}$				& $5.6484 \cdot 10^4$	& J/mol \\
Referential heat of crystallization				& $\Delta H_\text{ref}$	& $2.0920 \cdot 10^5$ 	& J/kg  \\
Ultimate degree of crystallization	& $\Phi_{\infty}$	& 0.5		& -				\\
Maximum crystallization rate			& $K_\text{max}$			& 1.64		& 1/s			\\
Temperature of maximum crystallization rate & $T_\text{max}$		& 423.15	& K				\\
Temperature half-width in crystallization rate					& $\Delta T$				& 80		& K				\\
Melt temperature at onset of crystallization					& $T^\lozenge$				& 538.15	& K				\\
Shear modulus						& $G$		& $1.1 \cdot 10^5$	& Pa	\\ 
Surface tension							& $\gamma$	& 0.036		& N/m			\\
\hline
\end{tabular}
\caption{Physical and rheological parameters for Nylon-66 melt, \cite{doufas_simulation_2000-1, shrikhande_modified_2006}. Note that no value for $\zeta=N_0l^2/3$ [m$^2$] is given in the literature. Since it does not play a role in the dimensionless model variants, we also do not specify it. }
\label{tab_nylon_physical_parameters}
\end{table}

\begin{table}[tb]
\centering
\begin{tabular}{|l l l |}
\hline
\multicolumn{3}{|l|}{\textbf{Model parameters for two-phase flow}}						\\
Description								& Symbol			& Value			\\
\hline
Giesekus mobility parameter								& $\alpha$			& 0.5 \\
Anisotropic drag coefficient	& $\sigma$		& 1.0				\\
Parameter for semi-crystalline relaxation time & $F$				& 20	 \\
Parameter for semi-crystalline relaxation time & &	\\
\qquad \quad for stress-driven model		& $\delta$		& 0.005	 \\
\qquad \quad for energy-driven model	& $\delta$		& 0.02	 \\
Parameter for flow-enhanced crystallization				& 	& 	\\
\qquad \quad for stress-driven model  &$\xi$ & 0.06\\
\qquad \quad for energy-driven model \quad &$\xi$ & 0.072\\
\hline
\end{tabular}
\caption{Model parameters associated to the two phases and the crystallization for melt spinning of Nylon-66, \cite{shrikhande_modified_2006}. }
\label{tab_shrikhande_model_parameters}
\end{table}

\begin{table}[tb]
\centering
\begin{tabular}{|l l l l|}
\hline
\multicolumn{4}{|l|}{\textbf{Process parameters}}											\\
Description								& Symbol			& Value		& Unit			\\
\hline
Fiber length							& $L$				& 1.6		& m				\\
Nozzle diameter							& $D_\text{in}$			& $2.159 \cdot 10^{-4}$ & m	\\
Temperature at inlet					& $T_\text{in}$			& 563.15	& K				\\
Velocity at inlet						& $v_\text{in}$			& 0.9272	& m/s			\\
Take-up velocity at outlet						& $v_\text{out}$			& 95		& m/s			\\
\hline
Air temperature							& $T_\mathrm{a}$			& 294.15	& K				\\
Air velocity, cross component			& $v_\mathrm{a}^\perp$				& 0.3048	& m/s			\\
Air velocity, tangential component \hspace*{1.5cm} \quad		& $v_\mathrm{a}^\parallel$				& 0			& m/s			\\
\hline
\end{tabular}
\caption{Process parameters for melt spinning of Nylon-66, (setup S01 in \cite{doufas_simulation_2000}).}
\label{tab_nylon_process_parameters}
\end{table}

The dimensionless Bingham function $B$ is related to the aerodynamic drag and can be interpreted as a local Bingham number,
\begin{equation*}
B(v_z,D,T) = 0.185 \left(\frac{\rho_\mathrm{a}(T)}{\mu_\mathrm{a}(T)} \,v_z D\right)^{0.39}.
\end{equation*}
The heat transfer coefficient $h$ is described by
\begin{equation*}
h(v_z,D,T) = 0.42 k_\mathrm{a}(T) \left(\frac{ \rho_\mathrm{a}(T)}{\mu_\mathrm{a}(T)}      \frac{v_z}{D^2}\right)^{1/3} \left(1 + \left(8\frac{v_\mathrm{a}^\perp}{v_z}\right)^2\right)^{1/6}
\end{equation*}
with cross component $v_\mathrm{a}^\perp$ of the air velocity, i.e., component that is perpendicular to the fiber. Both models stem from \cite{jeon_modeling_2008}, the last one is particularly based on the work of \cite{kase_matsuo_1965}.

In the considered melt spinning setup, velocity and temperature of the quench air are constant. For the specific values and the other process parameters used in the test case see Table \ref{tab_nylon_process_parameters}.

\section{Numerical treatment} \label{appendix_numerics}

\subsection{Collocation-continuation method for boundary value problems}\label{app:numerics}
Consider a boundary value problem of ordinary differential equations of the form
\begin{equation} \label{eq_bvp}
\ddz{\boldsymbol{y}} = \boldsymbol{f}(\boldsymbol{y}), \quad \quad \boldsymbol{g}(\boldsymbol{y}(0), \boldsymbol{y}(1)) = \mathbf{0}
\end{equation}
on the domain $I = (0,1)$, where $\boldsymbol{y}:I\rightarrow\mathbb{R}^N$ denotes the unknowns. The right hand side function $\boldsymbol{f}:\mathbb{R}^N \rightarrow \mathbb{R}^N$ and the boundary value function $\boldsymbol{g}:\mathbb{R}^N \times \mathbb{R}^N \rightarrow \mathbb{R}^N$ depend not only on the unknowns, but also on multiple parameters describing the dynamics of the underlying process. These parameters often make it difficult to solve the BVP without a good initial guess, which can be hard to find for given parameter settings. In this work we employ a collocation-continuation scheme that has been successfully used in various fiber formation problems, like glass wool production \cite{arne_fluid-fiber-interactions_2011}, melt blowing \cite{wieland_melt-blowing_2019}, dry spinning \cite{wieland_efficient_2019} or electrospinning \cite{arne_electrospinning_2017}.

\subsubsection*{Collocation scheme}
The collocation scheme is based on the three-stage Lobatto IIIa formula and can be interpreted as an implicit Runge-Kutta method of fourth order. It can be written as
\begin{equation*} \label{eq_collocation_method}
\begin{split}
&\boldsymbol{y}_{i+1} - \boldsymbol{y}_i - \frac{h_{i+1}}{6}\left(\boldsymbol{f}(\boldsymbol{y}_i) + 4 \boldsymbol{f}(\boldsymbol{y}_{i+1/2}) + \boldsymbol{f}(\boldsymbol{y}_{i+1})\right) = 0, \qquad \boldsymbol{g}(\boldsymbol{y}_0, \boldsymbol{y}_M) = \mathbf{0}, \\
&\textrm{with} \quad  \boldsymbol{y}_{i+1/2} = \frac{1}{2}(\boldsymbol{y}_{i+1} + \boldsymbol{y}_i) - \frac{h_{i+1}}{8} \left(\boldsymbol{f}(\boldsymbol{y}_{i+1}) - \boldsymbol{f}(\boldsymbol{y}_i) \right),
\end{split}
\end{equation*}
with collocation points $0=z_0<z_1<\ldots<z_M=1$, mesh size $h_i=z_i-z_{i-1}$ and the abbreviation $\boldsymbol{y}_i=\boldsymbol{y}(z_i)$, $i=0,\ldots,M$.
The resulting nonlinear system of $M+1$ equations is solved using a Newton method with analytically given or numerically computed Jacobian. For initialization, an initial mesh and an initial guess for the solution are required. The mesh is adapted by evaluating the residuals of the continuous solution to obtain a robust and effective algorithm. The residuals also serve as a stopping criterion to get an accurate approximation, \cite{Kierzenka_bvp4c_2001}. The collocation scheme is available in MATLAB as the routine \texttt{bvp4c.m}. We use the default settings.

\subsubsection*{Continuation procedure}
Since the performance of the collocation method crucially relies on the initial guess, we use a continuation approach and embed \eqref{eq_bvp} into a family of problems by introducing a continuation parameter vector $\bs{p} \in [0,1]^n$, $n \in \mathbb{N}$,
\begin{alignat*}{3}
&& \quad \ddz{\boldsymbol{y}} = \boldsymbol{\hat{f}}(\boldsymbol{y}; \boldsymbol{p}), & \qquad \boldsymbol{\hat{g}}(\boldsymbol{y}(0),\boldsymbol{y}(1); \boldsymbol{p}) = \mathbf{0}, & \\
\boldsymbol{\hat{f}}(\cdot ; \boldsymbol{1}) = \boldsymbol{f}, 	&& \quad \boldsymbol{\hat{g}}(\cdot , \cdot ; \boldsymbol{1}) = \boldsymbol{g},				& \qquad \boldsymbol{\hat{f}}(\cdot ; \boldsymbol{0}) = \boldsymbol{f}_0, \quad \quad \boldsymbol{\hat{g}}(\cdot , \cdot ; \boldsymbol{0}) = \boldsymbol{g}_0. &
\end{alignat*}
We choose a right hand side function $\boldsymbol{f}_0$ and a boundary value function $\boldsymbol{g}_0$ for which the analytical solution is known. This is the starting point to find a sequence of parameter vectors $\boldsymbol{0} = \boldsymbol{p}_0, \boldsymbol{p}_1, \boldsymbol{p}_2, \ldots, \boldsymbol{p}_l = \boldsymbol{1}$, where the solution for $\boldsymbol{p}_l$ corresponds to the desired solution of (\ref{eq_bvp}).
With this procedure we can enable or disable different contributions to the system as well as modify the boundary conditions along the continuation path. Success or failure of the continuation is determined by the choices of the step size and the path. 

Solving the BVP to $\boldsymbol{p}_{i+1}$ by using the solution belonging to $\boldsymbol{p}_i$ as initial guess is not alway possible such that step-size control is required. We use the approach of \cite{wieland_efficient_2019} which is briefly explained here for a one-dimensional parameter $p \in [0,1]$. Starting with an initial step size $\Delta p_0$, the BVP is solved twice by using one full step and two half steps, respectively. The step size is then adapted by comparing the resulting mesh sizes and the number of evaluations of the right hand side function. If the mesh size of the full step solution is greater than $k_1$-times the mesh size of the second half step, or if the number of evaluations in the full step is $k_2$-times more than the combined evaluations in both half steps, the step size is decreased by a factor $k_3$. Otherwise, if the full step evaluations are less than $k_4$-times the combined half step evaluations, the step size is increased by a factor $k_5$. Moreover, the steps size is also decreased by a factor $k_6$, if any call of the collocation routine fails. To handle continuation paths that do not contain existing intermediate solutions, the algorithm has a stopping criterion for too small step sizes, i.e., $\Delta p <\Delta p_\text{min}$. The parameters for all performed simulations are chosen as $\Delta p_0 = 0.1$, $\Delta p_\text{min} = 10^{-14}$, $k_1 =  1.1$, $k_2 = 0.9$, $k_3 = 1.5$, $k_4 = 0.7$, $k_5 = 1.5$, $k_6 = 10$.

The choice of the path is problem-specific, since there are mostly many ways to navigate through a parameter space and the existence of an intermediate solution is generally not known a priori. For this model class, we use the stress-free, non-crystallizing fiber with constant temperature and velocity as starting solution to the parameter vector $\bs{p} = \bs{0}$, i.e.,
\begin{alignat*}{7}
u &\equiv 1, \qquad\qquad	 & T &\equiv1, \qquad\qquad	& c_{zz} &\equiv1, \qquad\qquad & c_{rr} &\equiv1 \\
\omega &\equiv 0,  & S &\equiv0,				& x &\equiv 0, 	& a & \equiv 0.
\end{alignat*}
As continuation parameters we introduce  $p_\tn{Dr}$, $p_\delta \in [0,1]$ for draw ratio $\tn{Dr}$ and model parameter $\delta$. In particular, $\tn{Dr}$ and $\delta$ are replaced by $p_\tn{Dr}\tn{Dr} + (1-p_\tn{Dr})\tn{Dr}_0$ and $p_\delta \delta + (1-p_\delta)\delta_0$ with $\tn{Dr}_0 = 1$ and $\delta_0 = 0.05$. Moreover, to exclude air drag and gravitational effects we introduce $p_\tn{B}$, $p_\tn{Fr} \in [0,1]$ and replace the Bingham number $B$ by $p_\tn{B}B$ and the inverse Froude number $1/\tn{Fr}$ by $p_\tn{Fr}/\tn{Fr}$ in all equations. Finally, we include $p_T$, $p_S$, $p_x \in [0,1]$ to control the effect of heat transfer, phase orientation and crystallization. These parameters are added to the right hand side function of the differential equations for temperature, orientational tensor and crystallinity. This yields a seven-dimensional parameter space $\bs{p} = (p_\tn{Dr},p_T, p_\tn{Fr},p_\tn{B},p_x,p_S,p_\delta) \in [0,1]^7$ for the embedding of the energy-driven model (System~\ref{system_final_model_shrikhande}; EnMod) into a family of boundary value problems. For the interface boundary value problem (System \ref{system_final_model_doufas}; StMod) we even use two more continuation parameters, see Appendix~\ref{subsec:System2}.

To navigate  from $\bs{p} = \bs{0}$ to $\bs{p} = \bs{1}$  we perform a three-staged solution strategy for System~\ref{system_final_model_shrikhande}:
\begin{itemize}
\item[(A)] From $\bs{p} = \bs{0}$ to $\bs{p^A} = (1,1,1,1,0,0,0)$: \\
Increasing $p_{\tn{Dr}}$, $p_T$, $p_{\tn{Fr}}$ and $p_\tn{B}$ includes effects related to drawing, temperature, gravity and air drag.
\item[(B)] From $\bs{p^A}$ to $\bs{p^B} = (1,1,1,1,1,1,0)$: \\
Increasing $p_x$ and $p_S$ includes all effects related to crystallization.
\item[(C)] From $\bs{p^B}$ to $\bs{p} = \bs{1}$: \\
The model parameter $\delta$ is decreased to its correct value. 
\end{itemize}
In (A) and (B) we follow the diagonal path through the parameter space, thus $p_\tn{Dr} = p_T = p_\tn{Fr} = p_\tn{B}$ and $p_x = p_S$.
Step~(C) has numerical reasons since we observe a better convergence behavior of the approach using this procedure. At the end, after (C), we call the MATLAB routine \texttt{bvp4c.m} with relative error tolerance of $10^{-5}$ and absolute error tolerance of $10^{-10}$ to improve the accuracy of the solution.
For System~\ref{system_final_model_doufas} we apply a slightly different continuation path, see Appendix~\ref{subsec:System2}.

\subsection{Treatment of System~\ref{system_final_model_doufas}}\label{subsec:System2}
\subsubsection*{Interface transformation} System~\ref{system_final_model_doufas} is a boundary value problem with a free interface. For the numerical treatment it is convenient to transform it in the form of \eqref{eq_bvp}.  Hence, we introduce for $\tilde{z}\in[0,1]$
\begin{align*}
y^\mathrm{am}(\tilde{z})&=y(\tilde{z}\,z^\lozenge ), &&y\in\{v_z,T,c_{zz},c_{rr}\},\\
y^\mathrm{cr}(\tilde{z})&=y(\tilde{z}\,(z^\lozenge-1)+1), && y\in\{v_z,T,c_{zz},c_{rr},S,x,\omega\},
\end{align*}
where $y^\mathrm{am}$ and $y^\mathrm{cr}$ denote the unknown fiber functions before and after onset of crystallization, respectively. Then, the original boundary conditions are posed at $\tilde{z} = 0$ and the interface conditions at $\tilde{z} = 1$. This yields a boundary value problem with eleven equations on $[0,1]$. The unknown point $z^\lozenge$ of crystallization onset enters the equation system via the derivatives of the variables. It is determined by the additional condition $T^\mathrm{am}(1)=T^\mathrm{cr}(1) = T^\lozenge$.

\subsubsection*{Regularization}
In System \ref{system_final_model_doufas} multiple singularities arise if the crystallinity $x$ approaches one ($x\to 1$), which causes numerical difficulties. We use a regularization. In the right hand side of the system we replace the expressions
\begin{align*}
f(x)=\frac{1}{1-x} \qquad \text{by} \qquad \breve{f}(x)=\frac{1}{1-f_x(x)},
\end{align*}
where $f_x:[0,1] \rightarrow [0,x_\text{crit}]$, $x_\text{crit}<1$, is a continuously differentiable smoothing function
\begin{equation*}
f_x(x) =
\begin{cases}
x, 		& \hspace*{1.7cm} x < x_\text{crit}-\varepsilon\\
x+\frac{\varepsilon}{16}(-3+8\left(\frac{x_\text{crit}-x}{\varepsilon}\right)-6\left(\frac{x_\text{crit}-x}{\varepsilon}\right)^2+\left(\frac{x_\text{crit}-x}{\varepsilon}\right)^4),		& x_\text{crit}-\varepsilon \leq x < x_\text{crit}+\varepsilon\\
x_\text{crit}, &  x_\text{crit}+\varepsilon \leq x
\end{cases}
\end{equation*}
with critical crystallinity $x_\text{crit}$ and regularization parameter $\varepsilon$.
For the performed simulations, the parameters are $x_\text{crit}=0.997$ as in \cite{doufas_simulation_2000-1} and $\varepsilon=1-x_{crit}=0.003$.

\subsubsection*{Continuation strategy}
To handle the interface condition as well as the stress-dependent crystallization rate we 
apply a slightly different continuation strategy than described in Appendix~\ref{app:numerics}. We extend the parameter space by two additional continuation parameters $p_I$, $p_K$, yielding $\bs{p} = (p_\tn{Dr},p_T, p_\tn{Fr},p_\tn{B},p_x,p_S,p_\delta,p_I,p_K) \in [0,1]^9$.
The parameter $p_I$ controls the position of the interface.
As long as $p_T=0$, the considered fiber model is isothermal such that the interface condition cannot be satisfied and is hence replaced by $z^\lozenge = \hat{z}$ with $\hat{z}=0.1$.
When $p_T$ is increased to one, we pose $T(z^\lozenge) = p_I T^\lozenge + (1-p_I) \hat{T}$ where $\hat{T}=T(\hat z)$ for $p_I=0$.
To control the effect of flow-enhanced crystallization, we replace the crystallization rate $K$ by
\begin{align} \label{eqApp:conditionCrystRate}
\min(K,p_K K_1 + (1-p_K)K_2)
\end{align}
with $K_1=100$ and $K_2=10^{6}$.
To navigate from $\bs{p}=0$ to $\bs{p}=1$ we use the following strategy:\begin{itemize}
\item[(A)] From $\bs{p} = \bs{0}$ to $\bs{p^A} = (0,1,0,0,0,0,0,0,0)$: \\
Increasing $p_T$ includes temperature effects.
\item[(B)] From $\bs{p^A}$ to $\bs{p^B} = (0,1,0,0,0,0,0,1,0)$: \\
This step imposes the correct interface condition by increasing $p_I$.
\item[(C)] From $\bs{p^B}$ to $\bs{p^C} = (1,1,1,1,0,0,0,1,0)$: \\
Increasing $p_\tn{Dr}$, $p_\tn{Fr}$ and $p_\tn{B}$ includes the effects of drawing, gravity and air drag.
\item[(D)] From $\bs{p^C}$ to $\bs{p^D} = (1,1,1,1,1,1,1,1,0)$: \\
Increasing $p_x$ and $p_S$ includes crystallization. Furthermore, $\delta$ is changed to its actual value.
\item[(E)] From $\bs{p^D}$ to $\bs{p}=\bs{1}$: \\
Increasing $p_K$, all effects of flow-enhanced crystallization are incorporated.
 \end{itemize}
In (C) and (D) we follow the diagonal path through the parameter space, i.e., $p_\tn{Dr} = p_\tn{Fr} = p_\tn{B}$ and $p_x = p_S = p_\delta$.
After (E) we call the MATLAB routine \texttt{bvp4c.m} with relative error tolerance of $10^{-5}$, absolute error tolerance of $10^{-10}$ and without condition \eqref{eqApp:conditionCrystRate} to improve the accuracy of the solution and remove any restrictions on the crystallization rate.

\section*{Acknowledgment}
The support by the German BMBF (Project Vispi) is acknowledged.

\bibliographystyle{siam}
\bibliography{main}
\end{document}